\documentclass[journal,10pt,onecolumn]{IEEEtran}

\usepackage{amsmath}
\usepackage{amsfonts}
\usepackage{amssymb}
\usepackage{mathtools}
\usepackage{xifthen}
\usepackage{tikz}
\usepackage{tikzscale}		% scale tikz plot
\usepackage{tikz-3dplot} 
\usepackage{pgfplots}
\usepackage{acronym}
\pgfplotsset{compat=newest}
\usepackage[normalem]{ulem} 
\usepackage{caption}
\usepackage[labelformat=simple]{subcaption}
\usepackage{tabularx}
\usepackage{booktabs}

\usetikzlibrary{arrows.meta}
\usetikzlibrary{calc}
\usetikzlibrary{decorations.pathreplacing}
\usetikzlibrary{decorations.shapes}
\usetikzlibrary{fadings}
\usetikzlibrary{math}
\usetikzlibrary{plotmarks}
\usetikzlibrary{positioning}

\newcommand{\externalizeFigures}{false}
\ifthenelse{\equal{\externalizeFigures}{true}}
{
	\usepgfplotslibrary{external}
	\tikzexternalize[prefix=externalized/]
	\usepackage{shellesc}
}
{}

\definecolor{bpy1}{RGB}{255, 255, 122}
\definecolor{bpy2}{RGB}{253, 246, 82}
\definecolor{bpy3}{RGB}{250, 223, 82}
\definecolor{bpy4}{RGB}{246, 200, 87}
\definecolor{bpy5}{RGB}{243, 178, 98}
\definecolor{bpy6}{RGB}{241, 155, 114}
\definecolor{bpy7}{RGB}{238, 132, 133}
\definecolor{bpy8}{RGB}{237, 111, 153}
\definecolor{bpy9}{RGB}{236, 91, 175}
\definecolor{bpy10}{RGB}{201, 68, 196}
\definecolor{bpy11}{RGB}{165, 46, 218}
\definecolor{bpy12}{RGB}{131, 26, 240}
\definecolor{bpy13}{RGB}{96, 15, 245}
\definecolor{bpy14}{RGB}{61, 6, 245}
\definecolor{bpy15}{RGB}{26, 1, 245}
\definecolor{bpy16}{RGB}{0, 0, 234}
\definecolor{bpy17}{RGB}{0, 0, 188}
\definecolor{bpy18}{RGB}{0, 0, 142}
\definecolor{bpy19}{RGB}{0, 0, 95}
\definecolor{bpy20}{RGB}{0, 0, 49}

\definecolor{forestgreen}{RGB}{0, 220, 79}
\definecolor{Red}{RGB}{237, 27, 35}
\definecolor{LightGray}{gray}{0.9}

\acrodef{2d}[2D]{two-dimensional}
\acrodef{3d}[3D]{three-dimensional}
\acrodef{5g}[5G]{5th-generation}
\acrodef{awgn}[AWGN]{additive white Gaussian noise}
\acrodef{mpc}[MPC]{multipath component}
\acrodef{pdf}[PDF]{probability density function}
\acrodef{cdf}[CDF]{cummulative distribution function}
\acrodef{mmse}[MMSE]{minimum mean-square error}
\acrodef{simo}[SIMO]{single-input-multiple-output}
\acrodef{mimo}[MIMO]{multiple-input-multiple-output}
\acrodef{mmwave}[mmWave]{millimeter-wave}
\acrodef{mse}[MSE]{mean-square error}
\acrodef{ue}[UE]{mobile user}
\acrodef{los}[LoS]{line-of-sight}
\acrodef{nlos}[NLoS]{non-\ac{los}}
\acrodef{aoa}[AoA]{angle-of-arrival}
\acrodef{aod}[AoD]{angle-of-departure}
\acrodef{snr}[SNR]{signal-to-noise ratio}
\acrodef{dmc}[DMC]{dense multipath component}
\acrodef{fim}[FIM]{Fisher information matrix}
\acrodef{crlb}[CRLB]{Cram\'{e}r–Rao lower bound}
\acrodef{pcrlb}[PCRLB]{posterior \ac{crlb}}
\acrodef{eadf}[EADF]{effective aperture distribution function}
\acrodef{lhf}[LHF]{likelihood function}
\acrodef{rmse}[RMSE]{root mean square error} 
\acrodef{wssus}[WSSUS]{wide-sense stationary and uncorrelated scatterers}
\acrodef{usrp}[USRP]{Universal Software Radio Peripheral}
\acrodef{ofdm}[OFDM]{orthogonal frequency-division multiplexing}
\acrodef{tdma}[TDMA]{time-division multiple access}
\acrodef{agc}[AGC]{automatic gain control}
\acrodef{ctf}[CTF]{channel transfer function}
\acrodef{mrt}[MRT]{maximum ratio transmission}
\acrodef{lidar}[lidar]{light detection and ranging}
\acrodef{dpss}[DPSS]{discrete prolate spheroidal sequences}
\acrodef{pll}[PLL]{phase-locked loop}
\acrodef{slam}[SLAM]{simultaneous localization and mapping}
\acrodef{adc}[ADC]{analog to digital converter}
\acrodef{fpga}[FPGA]{field-programmable gate array}
\acrodef{pps}[1PPS]{1 pulse per second}
\acrodef{dpss}[DPSS]{discrete prolate spheroidal sequence}
\acrodef{snr}[SNR]{signal-to-noise ratio}
\acrodef{rf}[RF]{radio frequency}
\acrodef{icas}[ICAS]{integrated communication and sensing}
\acrodef{ota}[OTA]{over-the-air}
\acrodef{lo}[LO]{local oscillator}

\colorlet{border}{gray}

\usepackage{siunitx}
\DeclareSIUnit{\dBm}{dBm}
\catcode`\%=12\relax
\DeclareSIUnit[number-unit-product = ]\percent{\%}
\catcode`\%=14\relax
\newlength\figurewidth

\DeclareMathOperator*{\argmax}{arg\,max}

\DeclareMathOperator{\vect}{vec}
\DeclareMathOperator{\imagj}{j}

\newcommand{\Nfrequency}{N_\mathrm{f}}
\newcommand{\nfrequency}{n_f}

\usepackage{bm}			% for \bm ... bold math 

\begin{document}
	\title{Distributed MIMO Measurements for\\Integrated Communication and Sensing in an Industrial Environment}
	\author{%
		\IEEEauthorblockN{
			Christian Nelson\textsuperscript{$\dagger$},
			Xuhong Li\textsuperscript{$\dagger$},
			Aleksei Fedorov\textsuperscript{$\dagger$},
			Benjamin Deutschmann\textsuperscript{$\ddagger$},
			and Fredrik Tufvesson\textsuperscript{$\dagger$}
		}\\
		\IEEEauthorblockA{
			\textsuperscript{$\dagger$}Department of Electrical and Information Technology, Lund University, {22100} Lund, Sweden,\\
			\textsuperscript{$\ddagger$}Institute of {Communication} Networks and Satellite {Communications}, Graz University of Technology, 8010 Graz.
		}%
		\thanks{This research was partially funded by Connected Systems at Lund University, the strategic research area ELLIIT, and the REINDEER project of the European Union’s Horizon 2020 research and innovation program under grant agreement no. 101013425.}
	}
	\maketitle
	
\begin{abstract}Many concepts for future generations of wireless communication systems use coherent processing of signals from many distributed antennas. The aim is to improve communication reliability, capacity, and energy efficiency and provide possibilities for new applications through integrated communication and sensing. The large bandwidths available in the higher bands have inspired much work regarding sensing in the \ac{mmwave} and sub-THz bands; however, the sub-6\,GHz cellular bands will still be the main provider of wide cellular coverage due to the more favorable propagation conditions. In this paper, we present a measurement system and results of sub-6 GHz distributed \ac{mimo} measurements performed in an industrial environment.
	From the measurements, we evaluated the diversity for both large-scale and small-scale fading and characterized the link reliability. We also analyzed the possibility of multistatic sensing and positioning of users in the environment, {with the initial results showing a mean-square error below \SI{20}{\centi\metre} on the estimated position}. Further, the results clearly showed that new channel models are needed that are spatially consistent and deal with the nonstationary channel properties among the antennas.%
\end{abstract}
\section{Introduction}
\label{sec:Introduction}
With the advances of the fifth and sixth generation of mobile communication systems, new application fields are emerging such as vehicle-to-everything, machine-to-machine communication, and smart cities~\cite{3gppURLLC}. With new applications, requirements on, e.g., data rates, the number of connected devices and reliability increase. Furthermore, many applications need to be able to communicate and sense the environment. In order to do so \ac{icas}, where the same hardware and spectrum can be used for both purposes, has been given much attention in recent research. An application with very stringent requirements and that could benefit from \ac{icas} is the industrial scenario~\cite{Uusitalo2021}, which is the focus here.

As part of the development of wireless systems, new frequency bands are becoming available for communication, which also enables applications where sensing and communication coexist in the same band and use the same infrastructure and hardware~\cite{VanderPerre2019, Behravan2022}. However, due to the more favorable propagation conditions, most systems will probably still operate in the sub-6~GHz band. Furthermore, in fifth-generation networks, massive \ac{mimo} technology is seen as the main enabler of many requirements due to its potential to improve \ac{snr} and increase coverage due to the array gain, the ability to simultaneously serve many users, and improved reliability; the latter us partly due to the fact that small-scale fading effects effectively vanish as the number of antennas increases. 
This effect is called channel hardening~\cite{Willhammar2018, Willhammar2022}.

With the small-scale fading effects significantly reduced, the experienced reliability is to a large extent dependent on the large-scale fading effects. To combat this, distributed (massive) \ac{mimo}, where the antennas are spread over a larger physical area, has emerged as a candidate. Solutions such as cell-free massive MIMO \cite{Ngo2017, Bjoernson2020} and holographic MIMO \cite{Huang2020} are also being examined.
% Ta bort en eller flera alternativ om du vill
Another candidate is RadioWeaves~\cite{VanderPerre2019}, which is a proposed network architecture that combines distributed arrays and active, large intelligent surfaces~\cite{Hu2018} with distributed computation%u: Check edit accuracy
 to achieve high ultrareliable, low-latency communication. At the same time, the required amount of power to transmit is reduced due to the proximity of the users. 

Ultrareliable, low-latency communication is especially important in the industrial scenario. This scenario is a complex and rich environment from a propagation point of view, and channel characterization therefore becomes of great importance for designing radio systems to enable better communication quality and reliability. Hence, fading statistics need to be well studied in a given environment. 
%to allow the development or investigation of, e.g., network schemes and coding techniques~\cite{Oestges2011, tugratzEuCNC2023}.
These fading statistics are of great importance for the design of radio channel models and radio systems and for the development or investigation of, e.g., network schemes and coding techniques~\cite{Oestges2011, tugratzEuCNC2023} for a given~application.

With large aperture antenna arrays, such as in distributed \ac{mimo}, the commonly used assumption of \ac{wssus} channels is no longer valid.
%The commonly used assumption of \ac{wssus} channels is not valid for a large aperture antenna array.
%\Aleksei{reference, \cite{VanderPerre2019}?}.
There are two types of nonstationarities: (1) The first is related to the large aperture and distributed \ac{mimo}, in which the plane wave propagation assumption breaks down and becomes the spherical wave propagation assumption, i.e., an operation in the near field \cite{Aleksei1}. Different subarrays experience different channels, e.g., due to the various distances to a user and the difference in observing the \ac{los} and \ac{nlos} paths among different antennas. (2) The second is related to the temporal nonstationarity of the environment due to the fact that the channel statistics change over time in dynamic scenarios with the movement of users and other objects. 
If both of these are violated, the channel is said to be doubly underspread \cite{Matz2003b}.

% measurement equipment, requirements, other developed equipment
As new concepts emerge, there is also a need to test the feasibility of these concepts, and an important part of this is designing and building test beds. For distributed \ac{mimo}, different designs have been proposed \cite{Wassie2019, Zelenbaba2020, Stanko2021} and channel sounders and/or testbeds have been built. 
% Some of them are limited by cabling, others that they do not save all possible link combinations
% What they have in common is that they are all organized in a star network, which has the drawback that only the links between the antennas and the node can be measured and used for sensing, such as e.g. positioning and tracking of robots in an industrial setting.
In this work, we take this one step further and present a design of a truly distributed \ac{mimo} channel sounder organized as a mesh network where all the links between each antenna in the distributed array can also be measured and exploited for sensing purposes. 

% channel characterization and modeling (parameters, concepts investigated etc, relation to other peoples work)
With a measurement setup in place, channel measurements are needed to extract the relevant parameters for realistic channel models. In \cite{Loeschenbrand2019a, Zelenbaba2021, Guevara2021, Loeschenbrand2022, Nelson2023}, measurements of distributed channels were made. For the topic of joint communication and sensing, work has been done mainly in the mmWave bands in \cite{PrasobhSankar2021,Zhang2022a}.
Finally, theoretical work and simulations have been performed in \cite{fascista2023uplink} for a sub-6 GHz RadioWeave scenario for sensing~\cite{Fouda2022, Wymeersch2023}. Most measurements that have been conducted in terms of \ac{icas} have either been performed with a star-shaped design and/or for higher frequencies. Measurements with other topologies and/or sub-6~GHz frequencies are to a large extent lacking.

% industrial channels, \cite{Willhammar2022}

% positioning and tracking, delay, doppler, MUSIC, ESPRIT, ... 
% not sure where to place this paragraph yet, skip positioning and only focus on sensing as in above?

%\subsection{Related Work}

\subsection{Contributions}
In this paper, we describe a distributed \ac{mimo} channel sounder design. A whole new multi-ink measurement system has been developed to measure the dynamic properties of distributed antenna channels. As in all measurement setups, there is a need for calibration; here, we describe a practical implementation of how this can be done \ac{ota}, paving the way for even more advanced \ac{ota} calibration algorithms to be developed for more accurate system designs.
With this uniquely designed sounder, we conducted a measurement campaign in a realistic and dynamic industrial-like setting.
We analyzed the channel characteristics essential for reliability and nonstationarity aspects stemming from the large array and the dynamic environment. Finally, we exploited delay and/or Doppler information in order to explore the possibilities of sub-\SI{6}{\giga\hertz} channels for integrated communication and sensing in a mesh setup.
% think about how to clarify the contributions

\subsection{Structure of the Paper}
In Section~\ref{sec:radio-sig-model}, the signal model is presented.
Then, in Section~\ref{sec:meas-sys}, the developed measurement system is described.
{We describe the \ac{tdma} structure and describe why an \ac{agc} is implemented.}
In Section~\ref{sec:system-calibration}, {the need for system calibration is discussed as is how the sounder was calibrated in the presented measurement campaign. The measurement campaign is described in Section~\ref{sec:measurement-campaign}, along with the environment and the channel sounder configuration.}
The results are presented in Section~\ref{sec:analysis}, {including an analysis and discussion of both the communication aspect and sensing possibilities of distributed \ac{mimo}.} Finally, our conclusions {and future} work are outlined in Section~\ref{sec:conclusions}.

\subsection{Notation}
In this {paper}, $[\bm{a}]_i$ and $[\bm{A}]_{i,j}$ denote the $i$\textsuperscript{th} element of a vector $\bm{a}$ and the $(i,j)$\textsuperscript{th} entry of a matrix $\bm{A}$, respectively.  Estimated values are denoted with the hat symbol $\widehat{\cdot}$. The amplitude of a complex number $z$ is denoted by $|z|$, $z^*$ is the complex conjugate of $z$, and $\angle z$ is its phase. The Hadamarad product is denoted by $\odot$.
%The symbol $\bm{1}_{i \times j}$ denotes a $(i\times j)$-matrix of all ones.

%%%%%%%%%%%%%%%%%%%%%%%%%%%%%%%%%%%%%%%%%%
% Materials and Methods should be described with sufficient details to allow others to replicate and build on published results. Please note that publication of your manuscript implicates that you must make all materials, data, computer code, and protocols associated with the publication available to readers. Please disclose at the submission stage any restrictions on the availability of materials or information. New methods and protocols should be described in detail while well-established methods can be briefly described and appropriately cited.
% Give the name and version of any software used and make clear whether computer code used is available. Include any pre-registration codes.
%%%%%%%%%%%%%%%%%%%%%%%%%%%%%%%%%%%%%%%%%%
%%\section{Materials and Methods}
%
\section{Signal Model}\label{sec:radio-sig-model}
We consider $H_{\mathrm{a}}$ transceiver units distributed in the environment, and their positions are given as $\bm{p}_{n}^{(h)} = [p^{(h)}_{\mathrm{x},n},p^{(h)}_{\mathrm{y},n},{p}^{(h)}_{\mathrm{z},n}]^{\mathrm{T}} \in  \mathbb{R}^{3\times 1}$, with $h \in \mathcal{N}_{\mathrm{a}} \triangleq \left\{1,\dots, H_{\mathrm{a}}\right\}$.
In our setup, each transceiver unit supports two independent \ac{rf} chains, each connected to a single omnidirectional antenna.
It should be noted that a switched---possibly distributed---array can also be connected to the \ac{rf} chains for even larger setups. 
In the following signal model, we limit ourselves to the single antenna case for the sake of brevity in notation, but it can easily be extended to the switched array channel sounding system.
The $H_{\mathrm{a}}$th unit $\bm{p}_{n}^{{(H_a)}}$ represents the mobile agent, and the other units indexed by $h \in \left\{1,\dots, H_{\mathrm{a}}-1\right\}$ are the single antenna anchors at known positions.
At each time, the $h'$th unit acts as a transmitter and emits a periodic signal $\tilde{s}(t)$, and the other units $ \bm{p}_{n}^{(h)} $ with $h \in \mathcal{N}_{\mathrm{r}} \triangleq \left\{1,\dots, H_{\mathrm{a}}\right\}\setminus h'$ act as receivers.
Signals are represented by their complex envelopes with respect to a center frequency $f_{\mathrm{c}}$.
The signal received at the $h$ th antenna at the discrete observation time $n$ reads
\begin{equation}
\begin{aligned}
\bm{r}_{n}^{(hh')} &= \exp\left(\imagj2\pi \mu_{n}^{(hh')} t_{hh'} \right) \exp\left(-\imagj2\pi f_{\mathrm{c}} \epsilon^{(hh')}\right)
 \\ 
 \times &\sum_{l=1}^{L_n} \alpha_{l,n}^{(hh')} \exp\left(\imagj\eta^{(hh')}\right)\exp\left(-\imagj2\pi f_{\mathrm{c}} \tau_{l,n}^{(hh')}\right) \exp\left(\imagj2\pi \nu_{l,n}^{(hh')} t_{hh'}\right)  \bm{s} + \textbf{\emph{w}}_n^{(hh')},
\label{eq:SignalModel_FreqDiscrete} 
\end{aligned}
\end{equation}
where the first term comprises $L_{n}$ \acp{mpc}, $l \in \{1, \ldots , L_n \}$, with each being characterized by its complex amplitude $\alpha_{l,n}^{(hh')} \in \mathbb{C}$ and propagation delay $\tau_{l,n}^{(hh')}$.
Hardware impairments and imperfect synchronization are also characterized in the signal model.
More specifically, {$\mu_{n}^{(hh')}$ denotes the frequency offset between the $h$th unit and the $h'$th unit}, $ \eta^{(h)} $ denotes the unknown phase offset of the $h$th unit relative to a reference unit, $\epsilon^{(h)}$ denotes the time shift due to the clock offset of the $h$th unit, and $\nu_{l,n}^{(hh')}$ represents the Doppler shift at the time instant $t_{hh'}$ when the channel between the $h'$th transmit antenna and the $h$th receive antenna of the snapshot $n$ is measured.
Note that we omit the frequency dependency of the hardware impairment characteristics, given that a limited signal bandwidth of \SI{40}{\mega\hertz} is used.
Assuming we are transmitting on $\Nfrequency$ subcarriers, the vector $\bm{s} \in \mathbb{C}^{N_{\mathrm{f}}\times 1}$ accounts for the system response $\bm{g}\in \mathbb{C}^{N_{\mathrm{f}}\times 1}$ and the baseband signal spectrum $\bm{s}_{\mathrm{f}}\in \mathbb{C}^{N_{\mathrm{f}}\times 1}$; that is, $\bm{s} \triangleq \bm{g} \odot \bm{s}_{\mathrm{f}}$.
The system response is usually measured by a back-to-back calibration procedure.
The noise measurement processes $ w_n^{(h)}$ in \eqref{eq:SignalModel_FreqDiscrete} are independent \ac{awgn} with double-sided power spectral density {$N_{0}/2$}.

\section{Measurement System}\label{sec:meas-sys}
\begin{figure}
	\centering
	\includegraphics{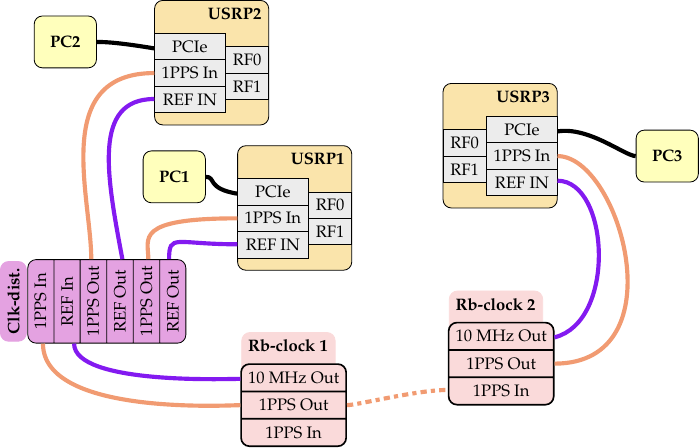}
	\caption{{An} illustration of a three-node, multilink setup. The dashed line between the two Rubidium clocks (Rb-clock 1 and Rb-clock 2) illustrates that if the two Rubidium clocks are well synchronized---over several hours---then they can be disconnected for some time without losing the synchronization of the radios. To the RF ports of the \acp{usrp}, one can either connect single antennas or switched arrays.\label{fig:measurement}}
\end{figure}

The multilink channel sounder has been developed utilizing the NI-\ac{usrp} {(National Instruments Corporation, Austin, TX, USA)} {and the software suite LabVIEW 2023}. The sounding system is portable and scalable, facilitating various measurement scenarios ranging from indoor and outdoor industrial settings to dense urban environments. The components of our multilink channel sounder system are listed in Table~\ref{tab:MultiLinkSystem} {and conceptual overview of the} \ac{rf} {parts are shown in Figure}~\ref{fig:measurement}.
\begin{table*}
    \caption{Hardware for the multilink measurement system.}\label{tab:MultiLinkSystem}
	\begin{tabular}{lcl}
        \toprule
        \textbf{{Hardware}} & \textbf{Amount} & \textbf{Description} \\
        \midrule
        NI-USRP 2953r \SI{40}{\mega\hertz}, {(National Instruments Corporation, Austin, TX, USA)} & 7 & \ac{usrp} \\
        SRS FS725, {(Stanford Research Systems Inc., Sunnyvale, CA, USA)} & 3 &\SI{10}{\mega\hertz} and 1PPS Rb standard \\
        SRS FS740, {(Stanford Research Systems Inc., Sunnyvale, CA, USA)} & 1 &\SI{10}{\mega\hertz} and 1PPS with GNSS \\
        Host computers & 7 & Radio control and logging data \\
        Hoverboard & 1& Acting as \ac{ue}\\
        Joymax SAF-6571RS3X antennas, {(Joymax Electronics Co., Ltd., Tao-yuan City, Taiwan)}& 13 & 12 as infrastructure and 1 on the \ac{ue} \\
        Ouster OSDome (128 lines), {(Ouster Inc., San Francisco, CA, USA)}& 1 & The lidar sensor\\
        Microstrain 3DM-GX5-25 (AHRS), {(Microstrain by HBK, Williston, VT, USA)}& 1 & 9-DoF IMU\\
        \bottomrule
    \end{tabular}%
\end{table*}

\begin{figure}
	\centering
\includegraphics{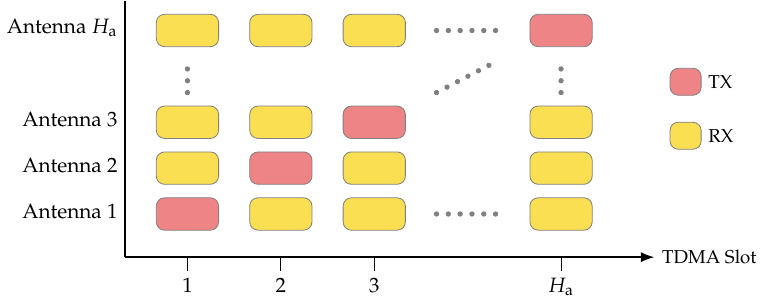}
\caption{During one \ac{tdma} slot, only one antenna is transmitting while all others are receiving. In the next TDMA slot, the next antenna is transmitting while all other are listening. By saving all channels, even the reciprocal ones, one can use the information for over-the-air calibration.\label{fig:tdma-time}}
\end{figure}

Designed for multilink channel sounding, the channel sounder records and stores all conceivable link combinations between antennas. To avoid interference among links, a \ac{tdma} strategy is employed.
Each antenna is assigned a unique time slot for signal transmission, during which the remaining antennas are set to receive mode. Figure~\ref{fig:tdma-time} provides a visual representation of this \ac{tdma} structure.
As a reference signal, the transmit unit uses a Zadoff--Chu sequence~\cite{Chu1972}.
The signal $\tilde{s}(t)$ is configured as an \ac{ofdm} symbol with Zadoff--Chu samples assigned to subcarriers \cite{Wassie2019}.
The sounding system also allows for the nullification of a specified number of carriers at the spectrum's edges, thus providing flexibility in bandwidth utilization.  
The channel sounder captures and streams the raw complex samples directly to the disk on the host computer for subsequent offline postprocessing, which may include symbol averaging.
Tight synchronization is necessary to achieve the \ac{tdma} structure. A one-pulse-per-second (1PPS) synchronization signal is distributed to all radios, as well as a stable \SI{10}{\mega\hertz} reference clock. Depending on the scenario, either synchronized Rubidium clocks or a GPS can be used to discipline the onboard clock and synchronize the triggers. For high-accuracy sensing measurements, rubidium clocks are the preferred choice{, we are using SRS FS725 and FS740 (Stanford Research Systems Inc., Sunnyvale, CA, USA)}.
%
%\vspace{-3pt}

We record all links, even those that could be assumed to be reciprocal.
By saving all channel transfer functions, we enable the possibility to evaluate \ac{ota} calibration algorithms.

\begin{figure}
\centering
\includegraphics{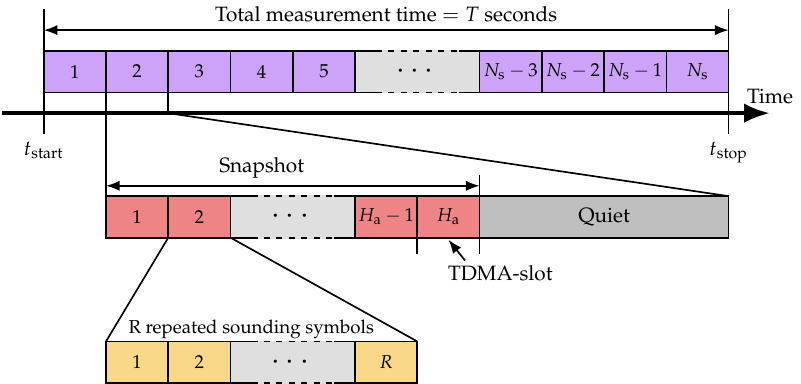}
\caption{The TDMA-based signal structure. {Each antenna is assigned a dedicated TDMA slot. During each transmission, the antenna transmits $R$ repetitions of the sounding signal, with some being used as guards and the rest for averaging to increase the signal-to-noise ratio.}\label{fig:tdma-structure}}
\end{figure}

The sounder is equipped with several adjustable parameters for the \ac{tdma} structure, as illustrated in Figure~\ref{fig:tdma-structure}. Initially, the reference symbol $\tilde{s}$, intended for transmission, is generated and stored in the \ac{fpga} memory of each \ac{usrp}. Subsequently, the number of repetitions of the reference symbol, denoted with $M$, is defined. $R \geq 2$, with the first symbol effectively serving as a cyclic prefix. Should \ac{agc} be used, a description of which will follow, $R \geq 3$. This setting accounts for the final symbol's potential distortion, as hardware adjustments may affect the receive gain during this period. Increasing the value of $M$ can improve the received \ac{snr} through symbol averaging. However, this improvement comes at the cost of extended transmission time and a reduced maximum resolvable Doppler frequency.

Furthermore, the structure includes $H_\mathrm{a}$ \ac{tdma} slots, where $H_\mathrm{a}$ corresponds to the number of antennas (see Figure~\ref{fig:tdma-time}). Following the activation and recording of all elements, the system can enter a quiet state for a duration of $B / \SI{120}{\mega\hertz}$ seconds, where $B$ represents the number of \ac{fpga} ticks and \SI{120}{\mega\hertz} is the \ac{fpga} clock rate.

\subsection*{Automatic Gain Control}
\begin{figure}
\centering
\includegraphics{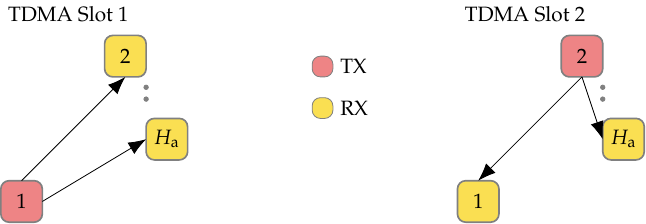}
\caption{Depiction of three of the $H_\mathrm{a}$ antennas distributed in space. In TDMA slot \num{1}, the antenna \num{1} is transmitting while all the others are listening. In then next \ac{tdma} slot, antenna $2$ is transmitting. Since antennas are distributed in space, it is clear from the figure that an \ac{agc} is needed; antenna $H_\mathrm{a}$ might need all the available gain when antenna $1$ is transmitting, while that same gain setting might saturate the \ac{adc} when antenna $2$ is transmitting.\label{fig:tdma_space}}
\end{figure}

Figure~\ref{fig:tdma_space} illustrates three of the $H_\mathrm{a}$ distributed single antennas. During the first \ac{tdma} slot, antenna 1 transmits while all other antennas are in receiving mode. Given the relative distance between antenna $1$ and antennas $2$ and $H_\mathrm{a}$, the latter may require maximization of its receive gain. In the subsequent \ac{tdma} slot, the next antenna in the sequence transmits and the rest assume receiving roles. In this \ac{tdma} slot, antenna $H_\mathrm{a}$, positioned closer to the transmitting antenna, might experience \ac{adc} saturation due to the preset gain of the receivers. This elementary example of a realistic scenario illustrates the need for an \ac{agc}. Due to the \ac{tdma} structure and how the antennas are distributed in space, the gain must be estimated and set within a couple of microseconds. Therefore, the \ac{agc} is implemented in the \ac{fpga} on board the radio to minimize latency. The implemented \ac{agc} is inspired by \cite{Sobaihi2012, Stanko2021}.

\section{System Calibration}\label{sec:system-calibration}
There are several system errors that need to be handled. Some of these errors are more pronounced than are others and stem from different sources, such as temperature variations, clock drift, clock offsets, and timing offset, to mention a few.
Many of these errors can be mitigated with well-synchronized measurement equipment and a stable temperature.
Now, we briefly describe five different errors and their possible sources. Let us start with the \emph{{time offset}}, which simply means that all the \acp{usrp} in the system must share the same notion of time so that all saved data from different antennas can be related to each other.
If the oscillators on the different transceivers do not provide the exact same carrier frequency, it will result in a \emph{{carrier frequency offset (CFO)}}.
If the aforementioned oscillator \acp{pll} locks in different phases, it will result in a \emph{{clock phase offset}}.
If the clock frequency of the \acp{adc} is different or imperfect, another frequency offset will be induced, namely, \emph{{sampling clock frequency offset}}.
Lastly, \acp{adc} might sample the signal at different times due to obtaining the trig signal at different times---e.g., due to different length cables---which is called \emph{{sampling time offset}}.
A summary can be found in Table~\ref{tab:systemerrors}.
\begin{table}
	\caption{Summary of measurement errors.}
	\label{tab:systemerrors}
	\begin{tabularx}{\textwidth}{ll} 
		\toprule
		\textbf{System Errors}& \textbf{Source} \\
  		\midrule
        Carrier Frequency Offset (CFO) & The oscillators do not provide the same frequency.\\
        Clock Phase Offset & The \acp{pll} lock on random---and different---phases.\\
        Sampling Clock frequency Offset&The clock frequency of the ADCs are not the same.\\
        Sampling Time Offset& The \acp{adc} samples at different times.\\
        Time Offset& The system does not share the same notion of time.\\
		\bottomrule
	\end{tabularx}%
\end{table}

To ensure accurate results, it is essential to perform a back-to-back calibration to remove as much as possible of the described errors.
During a back-to-back calibration, the cables from two \ac{rf} chains are connected together as close as possible to the antenna ports.
This gives the transfer function of the complete system between a pair of transceivers.
This step must be taken for all combinations of transceiver chains when all radios are operational with the settings intended for use during the measurement campaign.
This procedure enables the de-embedding of the radio channel from the measured \ac{ctf}, which, in addition to the propagation channel, also includes the influence of cables, connectors, the analog front-end, and digital processing.
Furthermore, if possible, it is recommended to characterize the antenna radiation pattern in an anechoic chamber to mitigate the effect of the antenna, thus isolating the wireless \ac{ctf}.
It is important to note that a hardware restart requires a recalibration. This requirement arises due to the reinitialization of the transmit-and-receive chains \ac{pll}, which lock onto a random phase after each restart.
If the purpose of the sounding is to measure metrics regarding the channel statistics for communication-related evaluation, then this requirement can be relaxed. However, for applications that require the use of coherent signals, such as accurate positioning algorithms, knowing the phase relations between all transceivers is crucial.

The next best thing is to perform a back-to-back calibration after a hardware reset using the same hardware (cables, connectors, etc.) as that used during the measurements.
With this method, there will be an error in the phase relationship since it is not possible to control at which phase the \acp{pll} of the different radios will lock.
If, for some reason, such as logistical constraints, any back-to-back calibration cannot not be performed, the use of an over-the-air calibration method in postprocessing is required.
This approach is feasible if the locations of the antennas are known.

In the following sections, we will describe the steps taken to achieve a calibrated data set. Due to logistical problems, we were unable to perform a back-to-back calibration, and hence we resorted to a combination of post-back-to-back calibration with two of the units to compensate for cable lengths and signal processing time. Then, we applied an over-the-air approach to compensate for the CFO and propagation delay. All steps require that at least a portion of the measured scenario is \emph{{static}} so that we can assume that the \ac{ctf} does not change during the calibration procedure. We ensured that we did not have any \emph{{time offset}} by syncing all computers to a local network time protocol (NTP) server. Then, we ensured that the \acp{fpga} shared the same notion of global time. We also assumed that there was no sampling clock frequency offset.

\subsection{DC Component}
This step is not, in the strict sense, a calibration procedure but is performed because our hardware uses direct down-conversion (DDC) from radio frequency down to baseband.
To avoid {\ac{lo}} leakage, the DC component is nulled by transmitting a 0 on the center subchannel ($f_\mathrm{c}$).
Hence, the (complex baseband) DC-component has to be interpolated by taking the average of the amplitude of the two neighboring complex coefficients and the average phase evolution as follows\footnote{$\mathrm{diff}\left(\left[a_1, a_2, \ldots, a_L\right]\right) = \left[ a_2 - a_1, a_3 - a_2, \ldots, a_L - a_{L-1}\right]$ is a {function that takes two consecutive values in the vector and their differences, with the resulting vector being one element smaller}; $\mathrm{unwrap}\left(\cdot\right)$ is the phase unwrapping function of Matlab}:

\begin{align}
    \widehat{\left[\bm{r}_n^{(hh')}\right]}_{f=0} = \frac{\left|\left[\bm{r}_n^{(hh')}\right]_{f=\Delta
 f}\right| + \left|\left[\bm{r}_n^{(hh')}\right]_{f=-\Delta
 f}\right|}{2} \cdot \exp{\left\{\mathrm{j} \left(\angle\left[\bm{r}_n^{(hh')}\right]_{f=-\Delta
 f} + \hat{\phi}_n\right)\right\}},
\end{align}
where $\hat{\phi} = \mathbf{E}\left\{\mathrm{diff}\left(\mathrm{unwrap}\left(\angle \bm{r}_n^{(hh')}\right)\right)\right\}$ is the average phase difference between two consecutive subcarriers after the phase has been unwrapped.

\subsection{Carrier Frequency Offset}
Even if a good reference clock is provided and distributed, there might be frequency drifts or offsets due, for example, to hardware impairments and/or temperature variations. 
To identify and remove possible carrier frequency offsets, we use a part of the measurement where \emph{all} antennas are static. If there is a carrier frequency offset, it comes from the oscillators and is not due to Doppler caused by movements.
Inspired by \cite{Tufvesson1999}, we identify carrier frequency offsets as follows. We collect the snapshots $\bm{r}^{(hh')}_n$ received at the antenna $h$ from the antenna $h'$ in the $\Nfrequency \times N_{st}$ matrix  $\bm{H}^{(hh')} = \left[\bm{r}^{(hh')}_0, \bm{r}^{(hh')}_1, \bm{r}^{(hh')}_2, \ldots, \bm{r}^{(hh')}_{N_{st}-1}\right]$, where $n \in \{0, 1, \ldots, N_\mathrm{st}-1\}$ are all static snapshots, and define the $N_{st} \times N_{st}$ shift matrix $\bm{S}$ \mbox{as follows:}

\begin{equation*}
    \bm{S} \triangleq \begin{pmatrix}
        0 & 0 & \cdots & 0 & 0\\
        1 & 0 & \cdots & 0 & 0\\ 
        0 & 1 & \cdots & 0 & 0\\
        \vdots & & \ddots & & \vdots\\
        0 & 0 & \cdots & 1 & 0
    \end{pmatrix}
\end{equation*}
{This is} applied $s$ times to shift the columns of $\bm{H}^{(hh')}$ as follows:
\begin{equation}
    \bm{C}^{(hh')} = \bm{H}^{(hh')} \odot \left(\bm{H}^{(hh')} \cdot \bm{S}^{s}\right)^{*}.
\end{equation}

{Discard} 
 the $s$ last columns of $\bm{C}^{(hh')}$ since they are all zeros 
\begin{equation}
    \bm{C}^{(hh')}_s \triangleq \left[\bm{C}^{(hh')}\right]_{1:\Nfrequency,1:N_{st}-s}.
\end{equation}

{The} average carrier frequency offset can now be estimated as follows:
\begin{equation}
    \hat{\mu}^{(hh')} = \angle \left( \sum_{n_f = 1}^{\Nfrequency}\sum_{n = 1}^{N_{st}-s}\left[\bm{C}^{(hh')}_s\right]_{\nfrequency,n} \right)
\end{equation}

{The correction} factor then becomes
\begin{equation}
    \exp{\left\{\mathrm{j}n\frac{\hat{\mu}^{(hh')}}{s}\right\}}, \qquad n \in \{0, 1, \ldots, N_{s}-1\} \, .
\end{equation}

\subsection{Delay Calibration}
Assuming line-of-sight conditions with no contributions from \acp{mpc}, the transfer function between antennas $h$ and $h'$ can be modeled as follows:

\begin{equation}
    \left[\bm{r}_n^{(hh')}\right]_{n_f} \approx \alpha_{l,n}^{(hh')}\exp{\left\{-\mathrm{j}2\pi f \frac{\|\bm{p}_n^{(h)} - \bm{p}_n^{(h')}\|}{\mathrm{c_0}} \right\}} \, ,
\end{equation}
where $\|\bm{p}_n^{(h)} - \bm{p}_n^{(h')}\| = d_n^{(hh')}$ denotes the scalar distance between antennas $h$ and $h'$.
Calibrating the delay $\tilde{\alpha}_{l,n}^{(hh')}$ can be omitted.
\begin{equation}\label{eq:cal-delay-line}
    \angle \left[\bm{r}_n^{(hh')}\right]_{n_f} = 2\pi f \cdot \frac{d_n^{(hh')}}{\mathrm{c_0}} = a\cdot f.
\end{equation}

{Equation} \eqref{eq:cal-delay-line} is a straight line with slope $a = 2\pi d_n^{(hh')}/\mathrm{c_0}$. Since both the frequencies and the constant distance during the snapshots selected for calibration are known, the 
``true'' slope is known. By estimating the measured slope, $\hat{a}$, of {Equation} \eqref{eq:cal-delay-line} and with the knowledge of the ground truth positions, a delay calibration can be formulated%AU: Check edit accuracy
 as follows:

\begin{equation}
    \hat{\epsilon}^{(hh')} \triangleq -\hat{a} +  2\pi d_n^{(hh')} / \mathrm{c_0}.
\end{equation}

{Of} course, this will not be true in practice, but this is a first approximation to enable calibration to compensate for delays induced by cables and signal processing on the \ac{fpga}.
If the channel is a non-line-of-sight condition, this approach will overestimate the delay and move the channel impulse response too far.

\section{Measurement Campaign}\label{sec:measurement-campaign}
\subsection{Environment}
\begin{figure}
   % \centering
    \subcaptionbox{\label{subfig:hall}}{\includegraphics[height=0.4\textwidth]{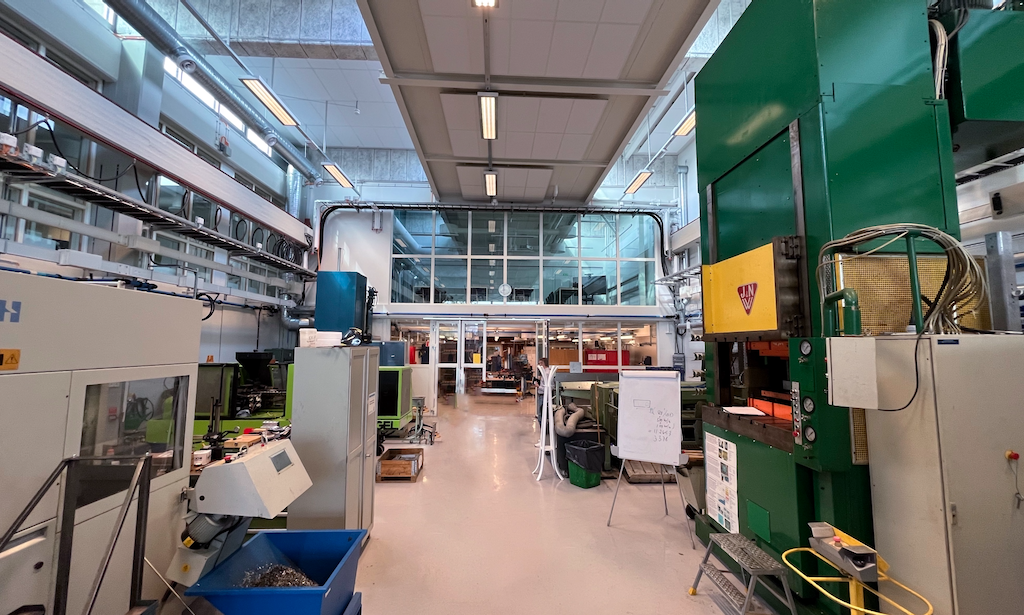}}%
    \hfill
    \subcaptionbox{\label{subfig:antennas}}{\includegraphics[height=0.4\textwidth]{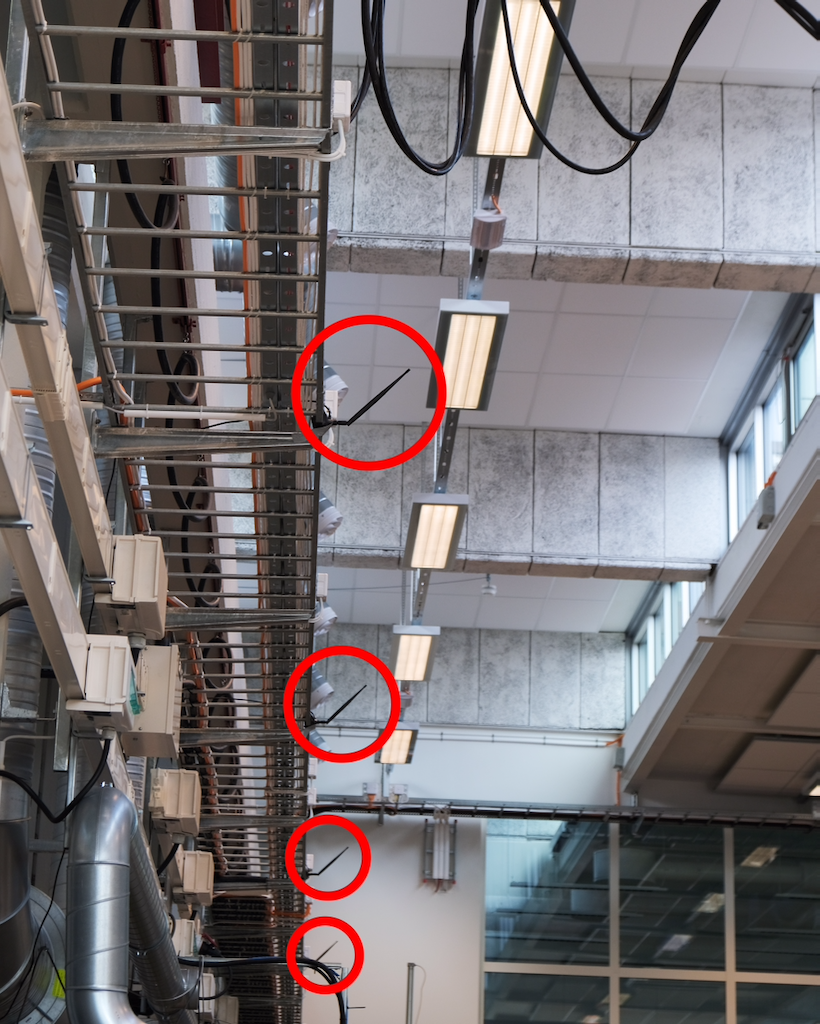}}%
    \caption{(\textbf{a}) {A photograph} %MDPI: We moved the position of the figure to avoid blank spaces, please confirm. %Author: confirmed.
 showing a view of the environment. The hall is approximately \qtyproduct{30 x 11}{\metre} with a ceiling height between \SIrange{8}{10}{\metre} depending on location. (\textbf{b}) {A photograph} showing the placement of four antennas {(circled in red)}. In total, there were twelve distributed antennas; six on each side of the hall. They were situated \SI{4}{\metre} above the floor, with a separation of \SI{4}{\metre}.}
    \label{fig:environment}
\end{figure}

The environment for our measurement campaign can be described by a typical industry hall for metal work, e.g., metal lathe, metal cutting. The dimensions are approximately \qtyproduct{30 x 11 x 8} {\metre} (L $\times$ W $\times$ H), see {Figure}~\ref{fig:environment}a. There are many metal objects and pieces of machinery that make for a rich wireless environment. Twelve static, {frequency} coherent, and distributed antennas were divided equally on each long side of the room, approximately \SI{4}{\metre} above the floor and separated by \SI{4}{\metre}; see Figure~\ref{fig:environment}b. {The infrastructure antennas were tilted approximately 45 degrees to obtain better coverage of the floor level area while strong reflections from the walls directly behind them were avoided. The free-space radiation pattern of the antennas is omnidirectional in cross section, but this will of course not be true as soon as it is attached to the metal structure and other objects close to its proximity. However, as previously mentioned in Section~\ref{sec:system-calibration}, we save and evaluate the \emph{{radio channel}}, which is the wireless propagation channel influenced by the antenna radiation pattern.} During the measurements, the facility was used as usual, with students working on projects and people moving around.

\subsection{Ground Truth}
To know where the channel samples are taken and to be able to quantify the accuracy of the radio-based position estimates, a \textit{{ground truth}} position is needed. This ground truth usually comes from high-quality global navigation satellite system (GNSS) signals when measurements are performed outdoors. In indoor scenarios, different approaches exist, e.g., camera-based motion capture, use of inertial measurement units (IMU), or a sensor fusion approach using cameras, lasers, and IMUs. However, the acquired ground truth positions must be at least an order of magnitude better than the estimates that are being evaluated.

\begin{figure}
\centering
\includegraphics[width=9.5cm]{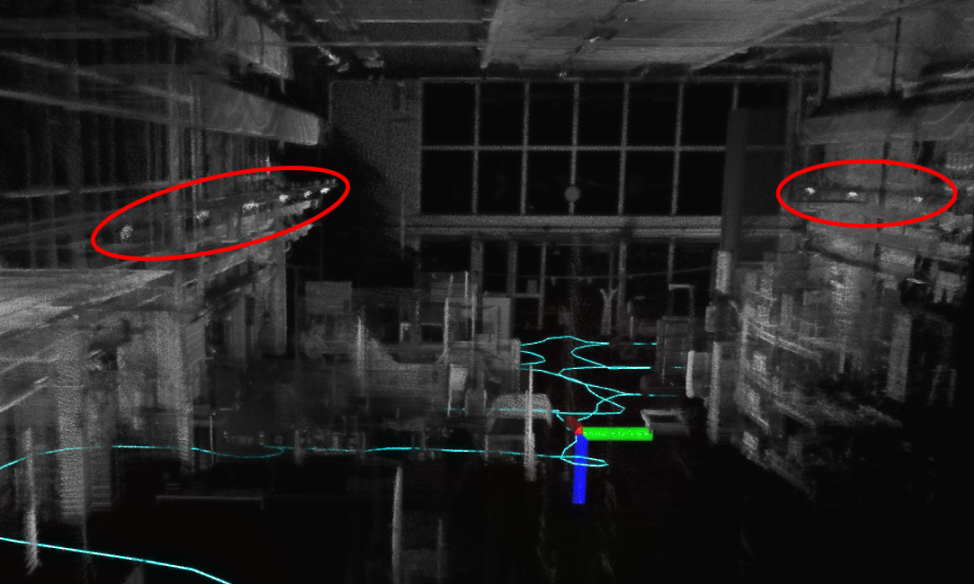}
\caption{{Example} output from FAST-LIO2. The figure depicts the intensity of the points in the scan, and the extra intensity spots on antenna locations with the help of reflective tape{, as seen in the circled regions. The cyan colored path shown is an example output of the ground truth position of the user.}\label{fig:lidar-scan}}
\end{figure}

In our case, all measurements were performed indoors, which ruled out a GNSS solution.
Therefore, a combination of a \ac{lidar} sensor {(Ouster OS-Dome 128, Ouster Inc., San Francisco, CA, USA)} and an IMU {(Microstrain 3DM-GX5-25 (AHRS), Microstrain by HBK, Williston, VT, USA)} was used to track the odometry of active and passive users.
The sensors were connected to a laptop running Ubuntu 20.04 and the Robot Operating System (ROS) \cite{quigley2009ros} Noetic.
All raw sensor messages were saved on disk to allow for the evaluation of different standard \ac{slam} algorithms. In this work, we used the open-source algorithm \emph{{FAST-LIO2}}; {see~\cite{fast_lio2} for details}. The sensors were mounted on a robot to track its position and orientation. During each measured scenario, a new map was constructed as the robot moved. Then, all maps were merged to obtain a single coordinate and reference system. At the mounting point on each of the distributed antennas, we put reflective tape to allow for the easier localization of the antennas on the map; see {Figure}%MDPI: There is no the citations of Figure 6 before the citation of Figure 7, please revise and ensure that the first citation of each figure appears in numerical order. %Author: Checked
~\ref{fig:lidar-scan}.

\subsection{Measured Scenarios}
\begin{figure}
\centering
\includegraphics[]{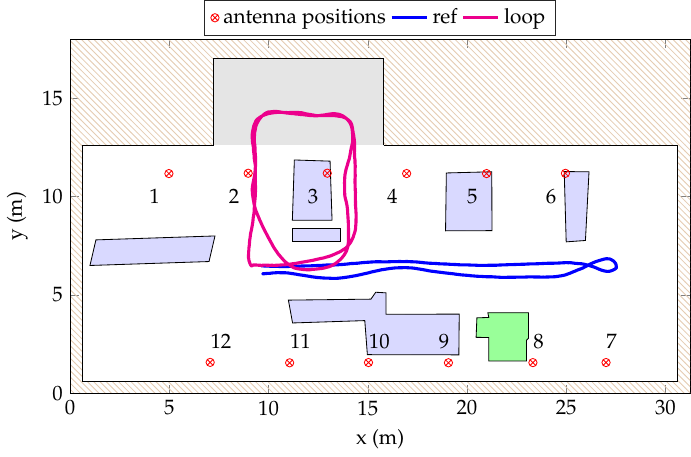}
\caption{{A} top-down schematic of the workshop where the measurements were performed. {The} \num{12}~static antennas are labeled according to the figure. The two paths are depicted where the samples were taken. {Machinery or equipment are marked purple, except} the machine colored in green which is tall enough to put antenna 8 in \ac{nlos} for the majority of the measurements. The part colored in gray is where the ceiling is much lower than the rest of the workshop.\label{fig:scen-groundtruth}}
\end{figure}

In this study, the moving agent traversed various paths to simulate conditions relevant to robotics or IoT devices in an industrial setting. Each path was traversed multiple times to ensure a robust statistical foundation for channel evaluations. This repetition also facilitates preliminary assessments of data-driven methodologies and machine-learning techniques based on different scenario realizations. All scenarios originated from a fixed position within the environment. For conciseness, this paper focuses on two primary scenarios, henceforth referred to as \emph{{ref}} and \emph{{loop}}; see Figure~\ref{fig:scen-groundtruth}, which collectively covers critical conditions such as \ac{nlos} and \ac{los} links.

In the \emph{{ref}} scenario, the robot navigates centrally through the room for approximately \SI{20}{\second}, executes a \SI{180}{\degree} rotation, and returns to its starting point. The whole sequence lasts around \SI{60}{\second}. In the \emph{{loop}} scenario, the robot drives two laps around some machinery and work tables. Parts of the trajectory have a much lower ceiling height than the rest of the hall. The \emph{{loop}} scenario lasts approximately \SI{80}{\second}. {For both scenarios the parameters of the channel sounder are detailed in Table}~\ref{tab:MultiLinkSysParams}.

\begin{table}
	\caption{{Channel} sounding parameters.}
	\label{tab:MultiLinkSysParams}
	\begin{tabularx}{\textwidth}{llll} 
		\toprule
		\textbf{Parameter Description}& \textbf{Value} & \textbf{Parameter Description}& \textbf{Value} \\
		\midrule
        Number of antennas, $H_\mathrm{a}$                     & \num{13}                      & Carrier frequency, $f_\mathrm{c}$       & \SI{3.75} {\giga\hertz}\\
		Frequency spacing, $\Delta f$           & \SI{78.125} {\kilo\hertz}      & Bandwidth, $BW$                         & \SI{40} {\mega\hertz}\\
        Active subcarriers, $N_\mathrm{f}$            & \num{449}                     & Number of subcarriers, $N_\mathrm{sc}$             & \num{512}\\
		Signal length, $\tau_\mathrm{max}$      & \SI{12.8} {\micro\second}      & Signal repetitions, $R$                 & \num{4}\\
		Snapshot length, $H_\mathrm{a}\cdot R\cdot\tau_\mathrm{max}$  & \SI{665.6}{\micro\second} & Repetition rate, $f_\mathrm{rep}$       & \SI{200}{\hertz} (\SI{5} {\milli\second})\\
		Max. resolvable velocity, $v_\mathrm{max}$  & \SI[per-mode=symbol]{8}{\metre\per\second} & Transmit power, $P_\mathrm{TX}$         & \SI{19} {\dBm}\\
        Measurement length, $T$                 & $T \in \left\{60, 80\right\}$ & Signal spacing, quite                   & \SI{4334.4} {\micro\second}\\
        Digital-to-analog back-off, $A_\mathrm{DAC}$ & \num{0.9} & &\\
		\bottomrule
	\end{tabularx}%
\end{table}
We used the remotely controlled robot as an active user in the environment, driving along different paths. The purpose of the measurement was to extract channel statistics for distributed \ac{mimo} and to collect channel samples to develop and evaluate positioning algorithms. To achieve this, we  collected channel data from routes in the somewhat more open space in the middle of the workshop as well as in the more obstructed parts with blocking from the machinery.

\section{Analysis and Discussion}\label{sec:analysis}

\subsection{Maximum Ratio Transmission}
To achieve reliable communication with low latency, i.e., no retransmissions, a high \ac{snr} is desired. 
If one has spatial diversity in the form of several transmit and/or receive antennas, as we have in our case, then it is shown that to maximize the \ac{snr} at time $n$, one should use the linear precoder presented in \cite{MRT1999}. Collect all uplink snapshots in the matrix
\begin{equation}
    \bm{H}_n = \left[\bm{r}_n^{(1,13)}, \ldots, \bm{r}_n^{(12,13)}\right],\quad\in\mathbb{C}^{(\Nfrequency \times H_\mathrm{a}-1)} \,
\end{equation}
then, using the $H_\mathrm{a}$ long column vector $\bm{\mathrm{e}} = [1, 1, \ldots, 1]^\mathrm{T}$ consisting of only ones
\begin{equation}
    \bm{H}_n^{\mathrm{MRT}} = \frac{\left(\bm{H}^*_n\odot\bm{H}_n\right)\bm{\mathrm{e}}}{\|\left(\bm{H}^*_n\odot\bm{H}_n\right)\bm{\mathrm{e}}\|}, \quad \in \mathbb{C}^{(\Nfrequency \times 1)} \,
\end{equation}
where the noise is assumed to be white Gaussian and uncorrelated with the signal.

\begin{figure}
\centering
\includegraphics[]{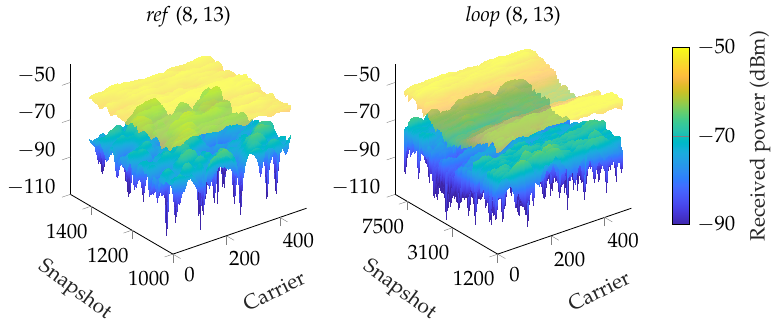}
\caption{{Using} \ac{mrt}, we clearly see a channel hardening effect where the deep fading dips are canceled. For visualization, a subset of the snapshots from the \emph{{ref}} scenario is selected, and only one, antenna 8 is plotted for comparison with \ac{mrt}. To the left is \ac{mrt} for a section of scenario \emph{{ref}}, and on the right, scenario \emph{{loop}} is shown.\label{fig:mrc-combined-surf}}
\end{figure}

\begin{figure}
\centering
\includegraphics[]{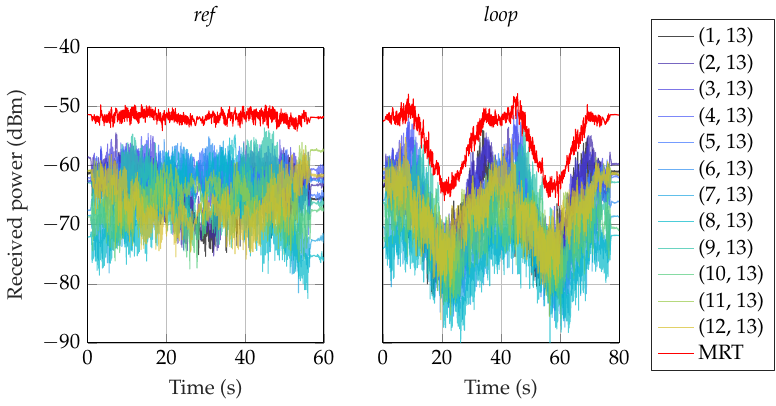}
\caption{{Using} \ac{mrt}, we clearly see a channel hardening effect where the deep fading dips are canceled. Here, all links are depicted using the average power of all subcarriers.\label{fig:summrc}}
\end{figure}

In Figure~\ref{fig:mrc-combined-surf}, two representative plots show the channel hardening effect from using distributed antennas; there are no more really deep fading dips. In the \emph{{ref}} scenario, we achieved an average array gain of \SI{13.8}{\dB}. In the \emph{{loop}} scenario, the average array gain was \SI{14.4}{\dB}. When we averaged all subcarriers, the results were similar, as shown in Figure~\ref{fig:summrc}. In the loop scenario, there were still variations in the received power levels in the order of \SI{10}{\dB}, and essentially all antennas experienced \ac{nlos} conditions at time \SI{20}{\second} and \SI{58}{\second}. However, despite the challenging propagation conditions, fading levels were small and the received power levels reasonably large. The key takeaway from these results is that there is much to gain from distributing antennas to combat small-scale {and} %MDPI: We removed the italic format, please confirm. %Author: Ok.
large-scale fading which enables reliable communication in challenging environments.

\subsection{Local Scattering Function}
In the case of a user moving in an industrial environment, the surroundings are usually cluttered with many objects that can have a considerable impact on the behavior of the propagation of wireless signals.
Hence, the fading process is nonstationary, which means that the wireless channel can be approximated by a piecewise stationary stochastic process where statistical parameters are valid locally (i.e., in small regions) \cite{Matz2003b}.
To extract parameters from nonstationary channels, we utilize the local scattering function defined in \cite{BernadoPhD, Matz2005}. 
This time-frequency-bounded function covers a stationarity region where the wireless channel can be well approximated by a \ac{wssus} process \cite{Matz2003a, Matz2003b}.

First, collect all snapshots between the antenna pair $(hh')$ in the matrix $\bm{H}^{(hh')}$

\begin{equation}
    \bm{H}^{(hh')} = \left[\bm{r}_1^{(hh')}, \ldots, \bm{r}_{N_\mathrm{ss}}^{(hh')}\right],\quad\in\mathbb{C}^{(\Nfrequency\times N_\mathrm{ss})}\, ,
\end{equation}

{Then,} 
 following the methodology outlined in \cite{BernadoPhD}, we denote the index of the sliding window in time and frequency with $k_t$ and $k_f$, respectively.  The size of the stationarity region is denoted with $M$ snapshots in the time domain and $N$ samples in the frequency domain.
Following \cite{Zelenbaba2021}, we use $N=\Nfrequency$ and henceforth drop $k_f$. The local scattering function is estimated as in \cite{BernadoPhD}:
\begin{equation}
    \left[\bm{C}^{(hh')}\right]_{k_t; n', p} = \frac{1}{IJ}\sum_{\mathrm{w}=0}^{IJ-1}\left| \left[\left(\bm{H}^{(hh')}\right)^\mathrm{w}\right]_{k_t; n', p}\right|^2
    \label{eq:LSF}
\end{equation}
where $n' \in \left\{0, \ldots, N-1\right\}$ is the delay index, and $p \in \left\{ -M/2, \ldots, M/2 - 1\right\}$ is the Doppler index. The local scattering function at $k_t$ represents the center value of the time-frequency region.
Let $\left(\bm{H}^{(hh')}\right)^\mathrm{w}$ be the windowed time-variant channel transfer function between each pair of antennas $h$ and $h'$ in the stationarity region $k_t$.
\begin{equation}
    \left[\left(\bm{H}^{(hh')}\right)^\mathrm{w}\right]_{k_t; n', p} = \sum_{m'=-M/2}^{M/2-1} \sum_{q'=-N/2}^{N/2-1} \left[\bm{H}^{(hh')}\right]_{m'-k_t,q'} \cdot \left[\bm{G}^\mathrm{w}\right]_{m',q'} \cdot \mathrm{e}^{-\imagj2\pi\left( \frac{pm'}{M}- \frac{n'q'}{N} \right)},
    \label{eq:tapered}
\end{equation}
where $m'$ and $q'$ are the relative time and frequency indices within each stationarity region.
The relationship between the absolute time index $n$ and the relative time index $m'$ is $n~=~\left(k_t - 1\right)\Delta_t + m' + M$ for $k_t \in \left\{1, \ldots, \lfloor\frac{N_{s}-M}{\Delta_t}\rfloor\right\}$, where $\Delta_t$ corresponds to the time shift between two consecutive regions of stationarity. 
The taper functions $\left[\bm{G}^\mathrm{w}\right]_{m',q'} = u_i\left[m' + M/2\right]\tilde{u}_j\left[q' + N/2\right]$ are the (separable) band-limited \acp{dpss} \cite{Slepian}, which are well localized within the region $\left[-M/2, M/2-1\right] \times \left[-N/2, N/2-1 \right]$. The sequences $u_i$ and $\tilde{u}_j$ are indexed by $i \in \left\{0, \ldots, I-1\right\}$ and $j \in \left\{0, \ldots, J-1\right\}$, respectively, with $\mathrm{w} = iJ + j$, which is the summation index in \mbox{{Equation} \eqref{eq:LSF}.}

For our measurements, we set $I=1$ and $J=2$ following the recommendations of \cite{BernadoPhD}. We choose $M = 75$, as the region of the minimum-time-stationarity region that corresponds to a duration of \SI{375}{\milli\second}. Considering the maximum speed of the mobile robot $v$ = \SI[per-mode=symbol]{1}{\metre\per\second}, the stationarity region becomes approximately $4.5$ wavelengths. As mentioned above, we choose $N=\Nfrequency$, assuming that the stationarity bandwidth is \SI{35}{\mega\hertz} since the relative bandwidth is less than \SI{1}{\percent}.
\subsection{Collinearity}\label{sec:collinear}
The collinearity metric between the local scattering function in two different time instances allows us investigate the extent of the stationarity region in time, $T_s\left[n\right]$; that is, how long the \ac{wssus} assumptions will hold \cite{BernadoPhD}. It should be noted that the stationarity time itself will be time dependent due to the changing environment. Stack the $N\times M$ elements of $\left[\bm{C}^{(hh')}\right]_{k_t; n', p}$ in a column vector $\bm{c}_{k_t}$ (without the superscript for readability) and define the collinearity $R\left[k_{t_1}, k_{t_2}\right]$ as follows:

\begin{equation}
R\left[k_{t_1}, k_{t_2}\right] = \frac{\bm{c}_{k_{t_1}}^\mathrm{T} \bm{c}_{k_{t_2}}}{\|\bm{c}_{k_{t_1}}\|\| \bm{c}_{k_{t_2}}\|}.
\end{equation}

\begin{figure}[htb]
    \centering
    \subcaptionbox{Threshold 0.9\label{subfig:collin-ref-3-th90}}{\includegraphics{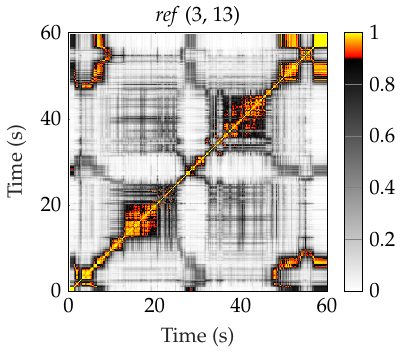}}
    \subcaptionbox{Threshold 0.7\label{subfig:collin-ref-3-th70}}{\includegraphics{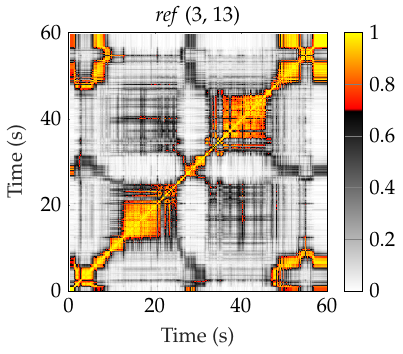}}\\ 
    \subcaptionbox{Threshold 0.9\label{subfig:collin-ref-9-th90}}{\includegraphics{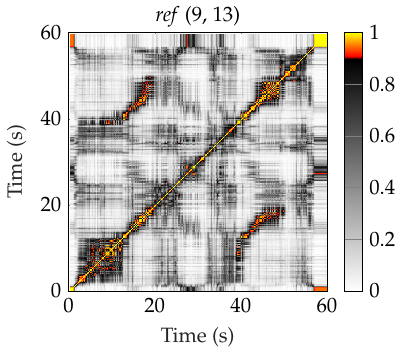}}
    \subcaptionbox{Threshold 0.7\label{subfig:collin-ref-9-th70}}{\includegraphics{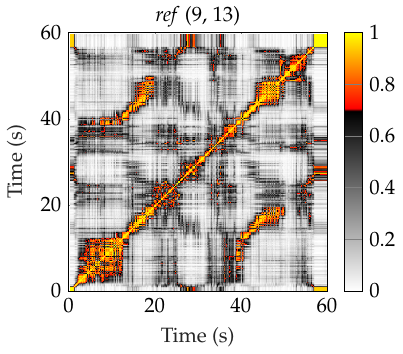}}
    \caption{The collinearity for two different links of the \emph{{ref}} scenario, with different threshold values. (\textbf{a}) Link {(3, 13)} with threshold 0.9, (\textbf{b}) link (3, 13) with threshold 0.7, (\textbf{c})  link (9, 13) with threshold 0.7, and (\textbf{d})  link (9, 13) with threshold 0.7.}
    \label{fig:collin-ref}
\end{figure}

\begin{figure}[htb]
    \centering
    \subcaptionbox{Threshold 0.9\label{subfig:collin-loop-3-th90}}{\includegraphics{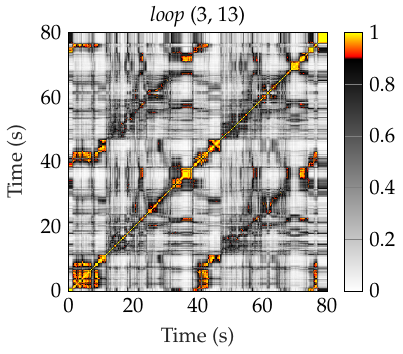}}
    \subcaptionbox{Threshold 0.7\label{subfig:collin-loop-3-th70}}{\includegraphics{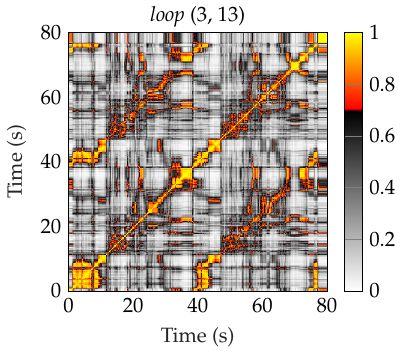}}\\
    \subcaptionbox{Threshold 0.9\label{subfig:collin-loop-9-th90}}{\includegraphics{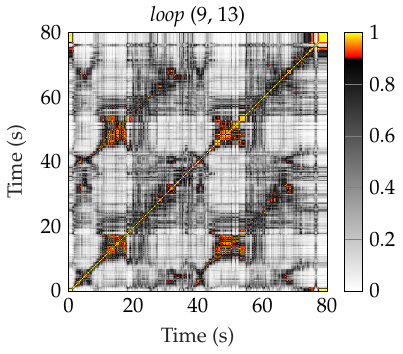}}
    \subcaptionbox{Threshold 0.7\label{subfig:collin-loop-9-th70}}{\includegraphics{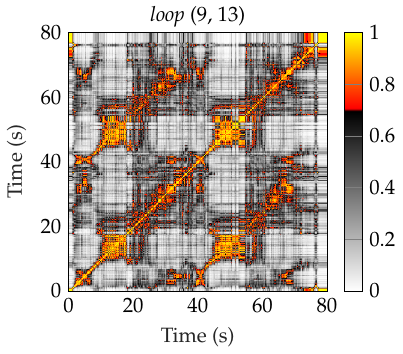}}
    \caption{The collinearity for two different links of the \emph{{loop}} scenario, with different threshold values. (\textbf{a}) Link (3, 13) with threshold 0.9, (\textbf{b}) link (3, 13) with threshold 0.7, (\textbf{c})  link (9, 13) with threshold 0.7, and (\textbf{d})  link (9, 13) with threshold 0.7}
    \label{fig:collin-loop}
\end{figure}

As in \cite{BernadoPhD},  we define the indicator function $\gamma\left[k'_t, \tilde{k'}_t\right]$ as
\begin{equation}
\gamma\left[k'_t, \tilde{k'}_t\right] =
    \begin{dcases*}
    1 & : $R\left[k'_{t}, \tilde{k'}_t\right] > c_\mathrm{th}$\,, \\[1ex]
    0 & : otherwise\,.
    \end{dcases*}
\end{equation}

where a threshold $c_\mathrm{th}$ is defined. From $\gamma$, the (time-varying) stationarity time, $T_s\left[n\right]$, can be estimated as the width of the region around the diagonal. In \cite{BernadoPhD}, the threshold $c_\mathrm{th}$ was somewhat randomly chosen as \num{0.9}. As seen in Figures~\ref{fig:collin-ref} and \ref{fig:collin-loop}, we select two (at random) links from the two scenarios \emph{{ref}} and \emph{{loop}}. We have also plotted how the regions would grow if $c_\mathrm{th} = 0.7$ instead.

In scenario \emph{{ref}}, the user was moving down in the middle of the workshop, then returning approximately the same path. In Figure~\ref{fig:collin-ref}a,b, we can see that on the way back, we move through a region where the time stationarity region is longer. Here, the channel statistics are valid for a longer distance. In Figure~\ref{fig:collin-ref}c,d,  it also looks like the off-diagonal regions indicate that we are actually moving through the same stationarity region on the way back since the collinearity between times \SI{15}{\second} and \SI{45}{\second} is above the threshold.

Performing a similar analysis on the collinearity plots of the \emph{{loop}} scenario, where the users performed two complete laps around some machinery, we can also indicate that we are in a stationarity region with similar statistics on the second lap. This becomes more apparent if we lower the threshold, $c_\mathrm{th}$, to \num{0.7}; see Figure~\ref{fig:collin-loop}. In general, the stationarity regions seem to become somewhat smaller because of the \ac{nlos} conditions.

\begin{figure}[htb]
\centering
\includegraphics{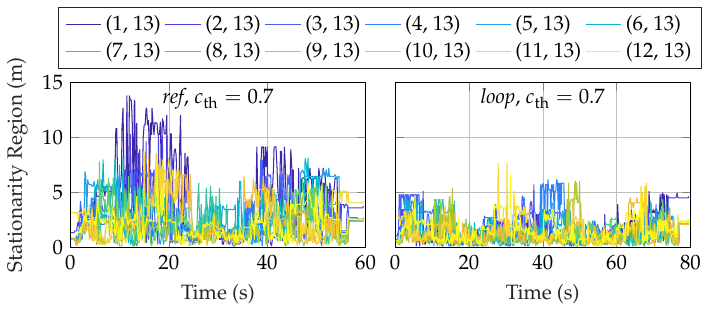}
\caption{The time-varying stationarity region (in m) with a threshold of \num{0.7}. \label{fig:tstat-th70}}
\end{figure}

\begin{figure}[htb]
\centering
\includegraphics{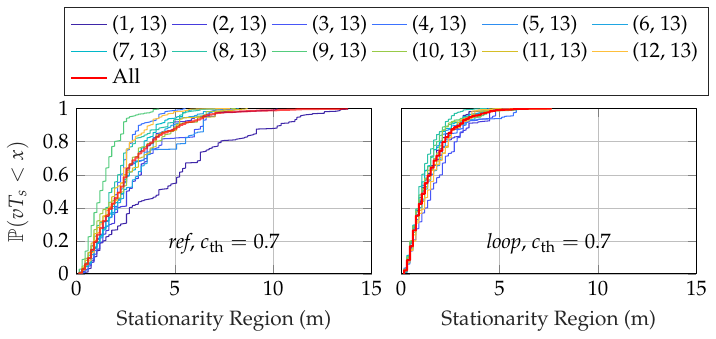}
\caption{The CDF of the time-varying stationarity region (in m) with a threshold of \num{0.7}. \label{fig:tstat-th70-ecdf}}
\end{figure}

In Figure~\ref{fig:tstat-th70}, we show the corresponding estimated time-varying stationarity regions in meters, $T_s\left[n\right]$, for the two scenarios when $c_\mathrm{th} = 0.7$. The median stationarity region is around \SI{2}{\metre}, see Figure~\ref{fig:tstat-th70-ecdf}, which means that the radio channel statistics vary while the UE is moving in the environment. Looking at the recorded statistics and the details of the environment, the rms delay spread, the Doppler power spectrum, and the LoS/NLoS states are changing for just a few meters of movement of the UE, hence the relatively short wide-sense stationarity regions.

\subsection{RMS Delay Spread}\label{sec:rms-delay-spread}
The power delay profile (PDP), $P_\tau$, can be calculated as the marginal expectation of the local scattering function {Equation} \eqref{eq:LSF} with respect to the Doppler domain as follows:
\begin{equation}
    \hat{P}_{\tau}[k_t; n'] = \frac{1}{M}\sum_{p}\left[\bm{C}^{(hh')}\right]_{k_t; n', p}
    \label{eq:PDP}
\end{equation}

\begin{figure}
\centering
\includegraphics{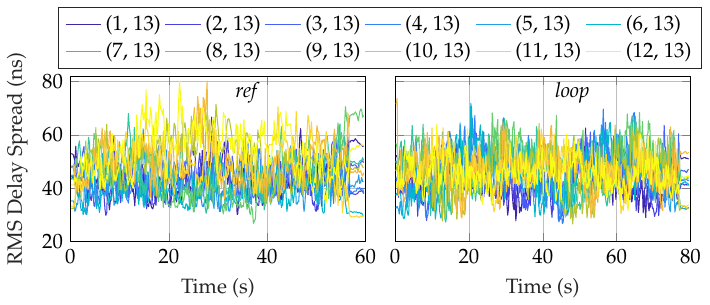}
\caption{RMS Delay spread for the two scenarios, calculated using the local scattering function. \label{fig:rmsdelayspread}}
\end{figure}

\begin{figure}
\centering
\includegraphics{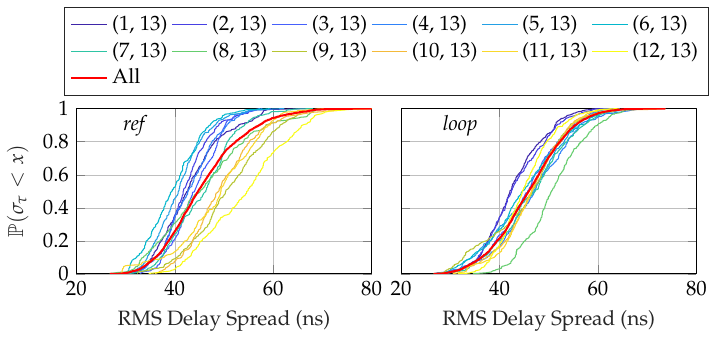}
\caption{The empirical CDF for the RMS delay spread for the two scenarios, calculated using the local scattering function. \label{fig:ecdf-rmsdelayspread}}
\end{figure}

From this, we can calculate the first and second moments $\tau$ and $\sigma_\tau$, respectively, as follows:
\begin{equation}\label{eq:rms-delay}
    \sigma_\tau \left[ k_t \right] = \sqrt{\frac{\sum_{n' = 0}^{N-1} (n'\tau_s)^2 \hat{P}_{\tau}[k_t; n']}{\sum_{n' = 0}^{N-1} \hat{P}_{\tau}[k_t; n']} - \tau\left[ k_t \right]^2},  \qquad \tau \left[ k_t \right] = \frac{\sum_{n' = 0}^{N-1} (n'\tau_s) \hat{P}_{\tau}[k_t; n']}{\sum_{n' = 0}^{N-1} \hat{P}_{\tau}[k_t; n']},
\end{equation}
where $\tau_s = 1/ (N\Delta f)$.
Figure~\ref{fig:rmsdelayspread} shows the rms delay spread for the different antennas over the two routes, and Figure \ref{fig:ecdf-rmsdelayspread} shows the corresponding CDFs. In calculating the moments in {Equation} \eqref{eq:rms-delay}, only contributions from the PDP that satisfied certain power thresholds were taken into account \cite{Czink2007}. The power threshold was selected as 5 dB above the noise floor to mitigate any spurious component, and 30 dB below the peak to only consider components that had a significant contribution.
The median rms delay spread was in the range \SI{38}{\nano\second} to \SI{54}{\nano\second}, with significant variations between both antennas and over the routes. We see that the results are in agreement with previous measurements in industry environments \cite{Rappaport1991, Karedal2004}, where the spread was also found to be around \SI{50}{\nano\second} in a similar-sized environment. In \cite{Holfeld2016}, machine-type communication between robot arms was measured in an industry environment. Measurements were made with a bandwidth of \SI{500}{\mega\hertz} and in a fixed position in the room due to the installation of the robot arm. In the their findings, the delay spread was somewhat lower, around \SI{30}{\nano\second}. Lastly, in \cite{Zhang2022b}, two wideband measurements were performed in what was classified as \emph{{indoor classroom}} and \emph{{industry}}. The dimensions of the room where the industrial measurements were taken were approximately half those of ours. They reported results of around \SI{70}{\nano\second} in both LOS and NLOS situations in the \emph{{industry}} scenario.

\subsection{Doppler Spectral Density}
An important metric to characterize dynamic channels is the Doppler spectral density (DSD). In Figure~\ref{fig:spec-music}, we present the time-variant DSD estimated with MUSIC \cite{Schmidt1986MUSIC} and ESPRIT \cite{Paulraj_1985ESPRIT}. Both methods are so-called super-resolution algorithms and both manage to estimate the Doppler well. There is a model parameter in both algorithms that must be estimate which is related to how many sources (tones) are expected, and this will vary in scenarios such as in the ones presented here. {Usually, the model order is estimated using, for example, the Akaike information criterion or minimum description length, but this study, we simply set the model parameter to two. The results showed that even in the challenging parts of the \emph{{loop}} scenario, both MUSIC and ESPRIT managed to find the dominant Doppler component.

\begin{figure}
\centering
\includegraphics{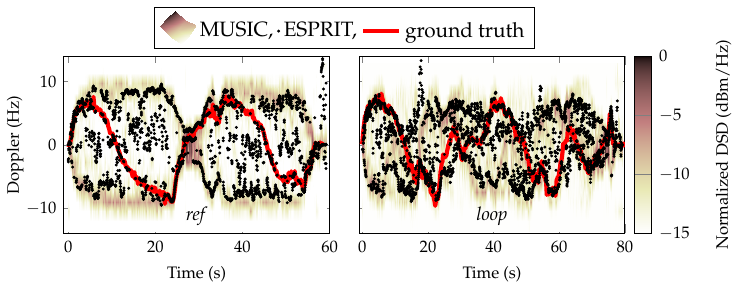}
\caption{The normalized Doppler spectral density (DSD) estimated with MUSIC for link (4, 13). The red line is the theoretical \ac{los} Doppler.\label{fig:spec-music}}
\end{figure}

\subsection{Doppler-Delay Positioning and Tracking}
To show how well our data set is suited for positioning and to hint at what accuracy can be achieved, we present the initial positioning and tracking results. 
We focus on an uplink positioning task where the agent transmits signals from which the D-\ac{mimo} infrastructure infers its position.
For this purpose, we focus on scenario \textit{{ref}}, where our aim is to track the {agent}~(\,\ref{pgf:agent}\,)
 at a ``true'' unknown position $\bm{p}_n^{(13)}$,
moving on a ({ground truth}) {trajectory}~(\ref{pgf:truth})
based on its uplink signals $\bm{r}_n^{(h\,13)}$ received by the D-\ac{mimo} {antennas}~(\,\ref{pgf:antennas-m}\,)
 at the ``true'' known positions $\{\bm{p}^{(h)} \ | \ 1 \leq h \leq 12\}$.
Before solving the positioning task, we first analyze the data.
For this purpose, we collect the available received snapshots until the current step $n$ along the trajectory into overlapping windows with a length of $N_\nu = 150$ and assemble them in matrices as follows:

\begin{align}
    \bm{H}^{(h13)}_{\tilde{n}} = \left[\bm{r}^{(h13)}_{n-N_\nu+1}, \ldots, \bm{r}^{(h13)}_{{n}}\right] \quad \in \mathbb{C}^{(\Nfrequency \times N_\nu)} \, ,
\end{align}

where we perform a rough delay calibration to account for the time shifts $\epsilon^{(h13)}$ introduced by the clock offsets of the receiving units $h$ w.r.t. the agent.
We formulate the $(\Nfrequency \times 1)$ temporal array response in the frequency domain through its elements as follows:

\begin{align}\label{eq:temporal-response}
    \left[\bm{b}\left(\bm{p}_n^{(13)}\right)\right]_{n_f} = \exp{\left(-\imagj2\pi f_{n_f} {\tau}_{n}^{(h13)}\right)}  \, ,
\end{align}

with $f_{n_f}$ denoting $n_f$th frequency bin of the signal in the complex baseband and ${\tau}_{n}^{(h13)}$ modeling the hypothetical propagation delay from the agent at $\bm{p}_n^{(13)}$ to the $h$th receiving antenna at $\bm{p}^{(h)}$.
We further formulate the $(N_\nu \times 1)$ Doppler array response in the time domain through its elements as follows:

\begin{align}\label{eq:Doppler-response}
    \left[\bm{c}\left(\bm{p}_n^{(13)},\bm{v}_n\right)\right]_{\tilde{n}} = \exp{\left(\imagj2\pi  t_{\tilde{n}} \nu_n^{(h13)} \right)}  \, ,
\end{align}

where $t_{\tilde{n}} \in \{0, \ldots , (N_\nu-1)/f_\mathrm{rep} \}$ corresponds to time instances of the current window of snapshots, and $\nu_n^{(h13)}$ models the hypothetical Doppler frequency shift depending on the agent position $\bm{p}_n^{(13)}$ and velocity $\bm{v}_n$ relative to the $h$th receiving antenna.
Note that we omit the dependence on \acp{mpc} $l>1$ in {Equations} \eqref{eq:temporal-response} and \eqref{eq:Doppler-response}. 
In our position and velocity estimator, we model \ac{los} propagation only where \ac{nlos} paths enter as disturbance.
Since the \ac{los} amplitudes are likely to be large compared to the \ac{nlos} amplitudes $\{{\alpha}_{l,n}^{(h13)} \ | \ l > 1\}$ and some of the receiving units $h$ will have the \ac{los} conditions (refer to Figures \ref{fig:environment} and \ref{fig:scen-groundtruth}), the D-\ac{mimo} units are likely to jointly agree on the true agent position, even in such an unfavorable industrial environment.
We compute the \textit{{nonphase-coherent}} empirical Bartlett spectrum {(}{for brevity, we omit the normalization term in the denominator of the classical Bartlett spectrum} from~{Equation} \eqref{eq:bartlett-derivation}{)} (see Appendix~\ref{sec:bartlett-derivation} for a derivation).

\begin{align}\label{eq:Bartlett}
    \hat{P}(\bm{p},\bm{v}) = \sum\limits_{h=1}^{12} \left| \bm{b}^\mathrm{H} \bm{H}^{(h13)}_{\tilde{n}}\bm{c}^* \right|^2
\end{align}
such that the contributions of all receiving antennas $h \in \{1 , \ldots , 12 \}$ are summed noncoherently as powers instead of complex-valued amplitudes because we do not have an accurate phase calibration available between our single-antenna receiving units. 
In the following, we assume that the agent is moving on a plane parallel to the $xy$-plane at a known height; hence, we aim for \ac{2d} positioning and velocity estimation in this work, well-knowing that~{Equation} \eqref{eq:Bartlett} is also suitable for \ac{3d} positioning.
We analyze the Bartlett spectrum around an observation step $n = 2544$ and hence employ a window of received signals $\tilde{n} \in \{2395,  \ldots, 2544\}$.

To evaluate the impact of only the temporal array response on the Bartlett spectrum, we choose $\bm{c}:=\bm{1}_{(N_\nu \times 1)}$ denoting a $N_\nu$-dimensional vector of all ones and evaluate \mbox{{Equation}~\eqref{eq:Bartlett},} which results in the spectrum depicted in Figure~\ref{fig:Bartlett-delay}. 
Due to the limited bandwidth of $BW = \SI{35}{\mega\hertz}$ and the respective delay resolution of approximately \SI{8.6}{\metre}, the resulting Bartlett spectrum is rather flat around the \textit{{true}} agent {position}~(\,\ref{pgf:agent}\,). 
Furthermore, imperfections in the delay calibration lead to a bias of the maximum $\argmax_{\bm{p}} \{\hat{P}(\bm{p})\}$ with respect to the true position $\bm{p}_{n}^{(13)}$. %

\begin{figure}
\centering
\setlength{\figurewidth}{0.7\columnwidth}
\input{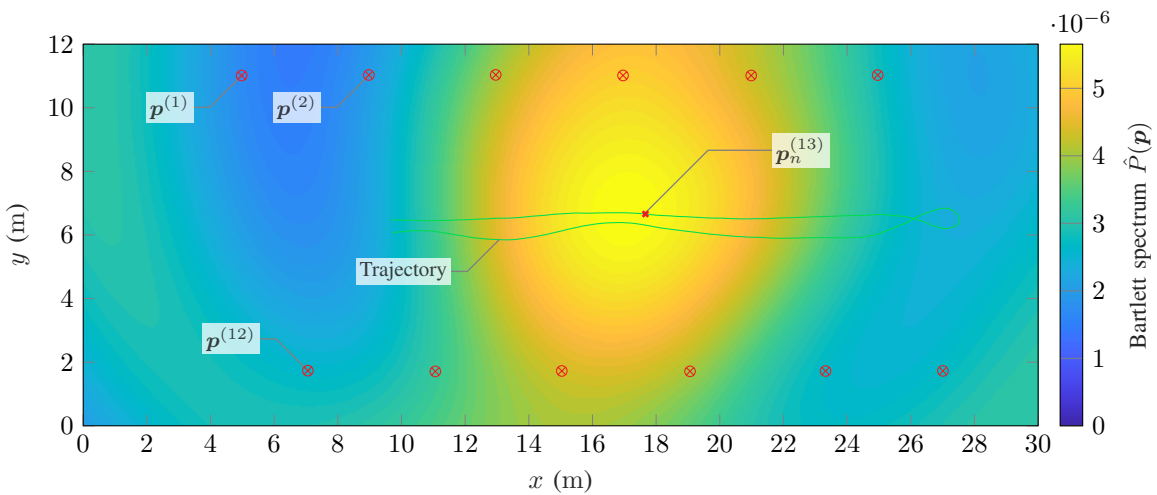}
\caption{{Bartlett} spectrum in the position domain exploiting only delay information.\label{fig:Bartlett-delay}}
\end{figure}

To evaluate the impact of only the Doppler array response on the Bartlett spectrum, we choose $\bm{b}:=\bm{1}_{(\Nfrequency \times 1)}$ and evaluate {Equation}~\eqref{eq:Bartlett}, resulting in the spectrum shown in Figure~\ref{fig:Bartlett-Doppler}. 
At the current time step $n$, the agent velocity is $\lVert \bm{v}_n \rVert \approx \SI{0.77}{\metre\per\second}$. 
At this speed, the Doppler array response is already much more informative (i.e., it exhibits a higher curvature) around the \textit{true} agent position than is the temporal array response at the chosen window size.
Hence, with a moving agent, the Doppler information quickly dominates over the delay information.

Another benefit of the Doppler array response in~{Equation} \eqref{eq:Doppler-response} is that it also depends on the agent velocity $\bm{v}_n$, and therefore $\argmax_{\bm{p},\bm{v}} \{\hat{P}(\bm{p},\bm{v}) \}$ is a joint position-velocity estimator.
Figure~\ref{fig:Bartlett_velocity} shows the resulting Bartlett spectrum in the velocity domain.
At the given speed, the Doppler array response likewise causes a distinct peak around the \textit{{true}} agent velocity~$\bm{v}_n$~(\,\ref{pgf:speed}\,).

\begin{figure}
\centering
\setlength{\figurewidth}{0.7\columnwidth}
\input{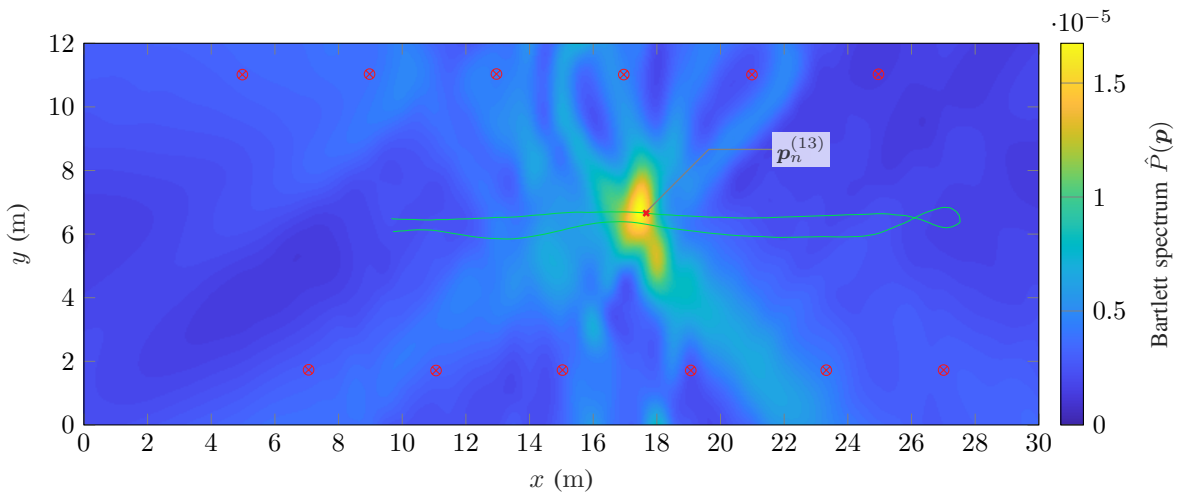}
\caption{{Bartlett} spectrum in the position domain exploiting only Doppler information.\label{fig:Bartlett-Doppler}}
\end{figure} %
\begin{figure}
\centering
\setlength{\figurewidth}{0.4\columnwidth}
% This file was created by matlab2tikz.
%
%The latest updates can be retrieved from
%  http://www.mathworks.com/matlabcentral/fileexchange/22022-matlab2tikz-matlab2tikz
%where you can also make suggestions and rate matlab2tikz.
%
\begin{tikzpicture}

\begin{axis}[%
width=\figurewidth,
height=\figurewidth,
at={(0\figurewidth,0\figurewidth)},
scale only axis,
point meta min=0,
point meta max= 2.3279e-05,
axis on top,
xmin=-1,
xmax=1,
xlabel style={font=\color{white!15!black}},
xlabel={$v_x$ (\SI{}{\metre\per\second})},
ymin=-1,
ymax=1,
ylabel style={font=\color{white!15!black}},
ylabel={$v_y$ (\SI{}{\metre\per\second})},
axis background/.style={fill=white},
colormap={mymap}{[1pt] rgb(0pt)=(0.2422,0.1504,0.6603); rgb(1pt)=(0.25039,0.164995,0.707614); rgb(2pt)=(0.257771,0.181781,0.751138); rgb(3pt)=(0.264729,0.197757,0.795214); rgb(4pt)=(0.270648,0.214676,0.836371); rgb(5pt)=(0.275114,0.234238,0.870986); rgb(6pt)=(0.2783,0.255871,0.899071); rgb(7pt)=(0.280333,0.278233,0.9221); rgb(8pt)=(0.281338,0.300595,0.941376); rgb(9pt)=(0.281014,0.322757,0.957886); rgb(10pt)=(0.279467,0.344671,0.971676); rgb(11pt)=(0.275971,0.366681,0.982905); rgb(12pt)=(0.269914,0.3892,0.9906); rgb(13pt)=(0.260243,0.412329,0.995157); rgb(14pt)=(0.244033,0.435833,0.998833); rgb(15pt)=(0.220643,0.460257,0.997286); rgb(16pt)=(0.196333,0.484719,0.989152); rgb(17pt)=(0.183405,0.507371,0.979795); rgb(18pt)=(0.178643,0.528857,0.968157); rgb(19pt)=(0.176438,0.549905,0.952019); rgb(20pt)=(0.168743,0.570262,0.935871); rgb(21pt)=(0.154,0.5902,0.9218); rgb(22pt)=(0.146029,0.609119,0.907857); rgb(23pt)=(0.138024,0.627629,0.89729); rgb(24pt)=(0.124814,0.645929,0.888343); rgb(25pt)=(0.111252,0.6635,0.876314); rgb(26pt)=(0.0952095,0.679829,0.859781); rgb(27pt)=(0.0688714,0.694771,0.839357); rgb(28pt)=(0.0296667,0.708167,0.816333); rgb(29pt)=(0.00357143,0.720267,0.7917); rgb(30pt)=(0.00665714,0.731214,0.766014); rgb(31pt)=(0.0433286,0.741095,0.73941); rgb(32pt)=(0.0963952,0.75,0.712038); rgb(33pt)=(0.140771,0.7584,0.684157); rgb(34pt)=(0.1717,0.766962,0.655443); rgb(35pt)=(0.193767,0.775767,0.6251); rgb(36pt)=(0.216086,0.7843,0.5923); rgb(37pt)=(0.246957,0.791795,0.556743); rgb(38pt)=(0.290614,0.79729,0.518829); rgb(39pt)=(0.340643,0.8008,0.478857); rgb(40pt)=(0.3909,0.802871,0.435448); rgb(41pt)=(0.445629,0.802419,0.390919); rgb(42pt)=(0.5044,0.7993,0.348); rgb(43pt)=(0.561562,0.794233,0.304481); rgb(44pt)=(0.617395,0.787619,0.261238); rgb(45pt)=(0.671986,0.779271,0.2227); rgb(46pt)=(0.7242,0.769843,0.191029); rgb(47pt)=(0.773833,0.759805,0.16461); rgb(48pt)=(0.820314,0.749814,0.153529); rgb(49pt)=(0.863433,0.7406,0.159633); rgb(50pt)=(0.903543,0.733029,0.177414); rgb(51pt)=(0.939257,0.728786,0.209957); rgb(52pt)=(0.972757,0.729771,0.239443); rgb(53pt)=(0.995648,0.743371,0.237148); rgb(54pt)=(0.996986,0.765857,0.219943); rgb(55pt)=(0.995205,0.789252,0.202762); rgb(56pt)=(0.9892,0.813567,0.188533); rgb(57pt)=(0.978629,0.838629,0.176557); rgb(58pt)=(0.967648,0.8639,0.16429); rgb(59pt)=(0.96101,0.889019,0.153676); rgb(60pt)=(0.959671,0.913457,0.142257); rgb(61pt)=(0.962795,0.937338,0.12651); rgb(62pt)=(0.969114,0.960629,0.106362); rgb(63pt)=(0.9769,0.9839,0.0805)},
colorbar,
colorbar style={ylabel style={font=\color{white!15!black}},
y tick scale label style={xshift=0.6cm}, 
xshift = -0.3cm, %
width=0.3cm,%
ylabel={\small Bartlett spectrum $\hat{P}(%\bm{H},%^{(h1)}
\bm{v})$}
}
]
\addplot [forget plot] graphics [xmin=-1.00244498777506, xmax=1.00244498777506, ymin=-1.00244498777506, ymax=1.00244498777506] {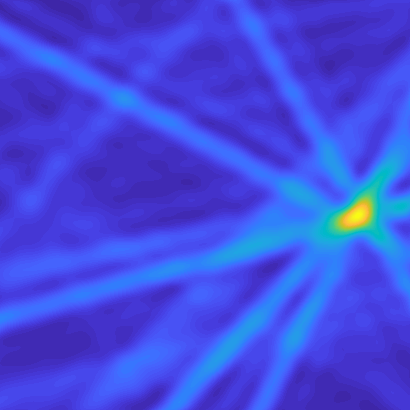};
\addplot [color=Red, line width=0.5pt, only marks, mark size=1.5pt, mark=triangle, mark options={solid, Red}, forget plot]
  table[row sep=crcr]{%
0.767682128696446	-0.068348192291201\\
};\label{pgf:speed}

\draw[color=gray, line width=0.5pt]    % agent
(0.7677,-0.0684) -- %
(0.5677,-0.2684) -- %
(0.3677,-0.2684); %
\node[left, align=right ,font={\footnotesize},fill=white,
opacity=0.75,inner sep=0.5mm, xshift=0.0mm] at %
(axis cs:0.3677,-0.2684){%\scalebox{0.9}
{$\bm{v}_{n}$}};

\end{axis}
\end{tikzpicture}%
\caption{{Bartlett} spectrum in the velocity domain exploiting only Doppler information.\label{fig:Bartlett_velocity}}
\end{figure}
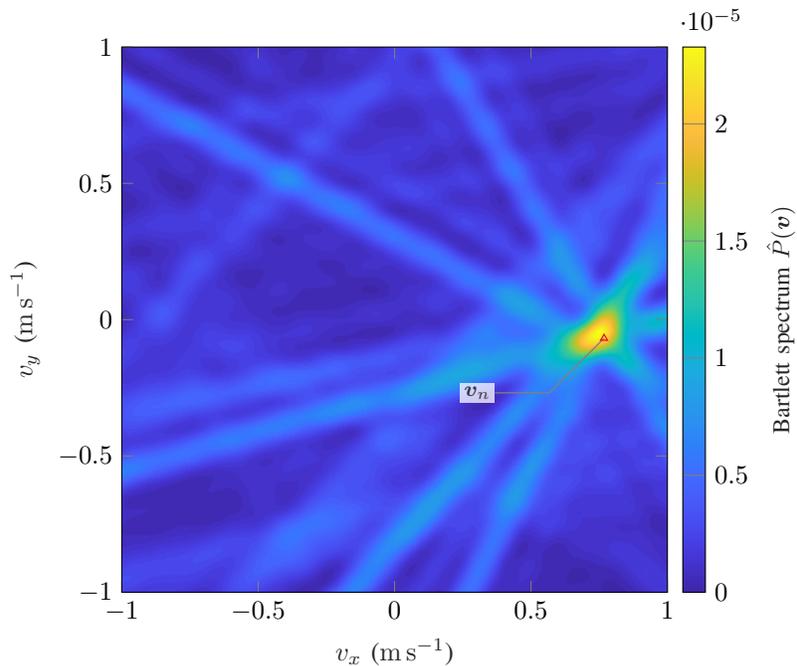

% %%%%%%%%%% %%%%%%%%%% %%%%%%%%%% %%%%%%%%%% %%%%%%%%%%
Although reasonable position and velocity estimates can be extracted from a single snapshot of data, state filtering over all snapshots along the trajectory achieves much better~results.

To showcase initial results, we employ the empirical Bartlett spectrum from Equation~\eqref{eq:Bartlett} in a particle filter together with a nearly constant velocity state-space model (cf.\,\cite{BarShalom2002Estimation} p.\,274).
Figure~\ref{fig:Batlett_Traj} shows the resulting estimates~(\ref{pgf:est}) in the position domain compared to the ground truth trajectory.~(\ref{pgf:truth}). 
It is observable that the estimates slowly converge to the true trajectory and follow it closely until the agent performs its \SI{180}{\degree} turn. 
In the curve, the velocity decreases, as does the information from the Doppler array response in the position domain because the \textit{{sensitivity}} of a Doppler frequency shift with respect to the position; i.e., $\partial \nu_n^{(h13)} /\partial \bm{p}_n^{(13)}$ depends linearly on the (radial) velocity of the agent relative to the $h$th unit.
The estimation accuracy decreases for a moment until the agent moves at maximum speed and the position estimates converge again. Using the Bartlett beamformer-based implementation, we achieve a positioning \ac{mse} of  \SI{18.4}{\centi\metre} with respect to our ground truth. 
These initial results highlight the potential of the dataset for positioning and tracking and set the stage for future work on more elaborate estimators.

\begin{figure}
\centering
\setlength{\figurewidth}{0.8\columnwidth}
\input{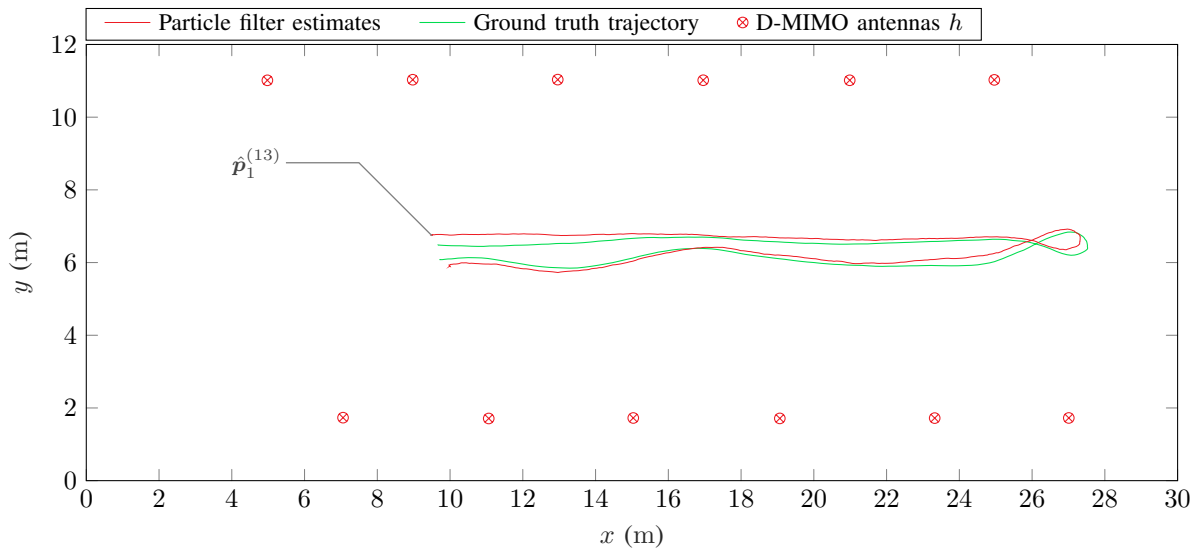}
\caption{Initial trajectory estimation result based on the Bartlett estimator using~{Equation} \eqref{eq:Bartlett} with a particle filter. The particle filter is initialized at a position $\bm{p} = [9.5,6.5]^\mathrm{T}$, close to its first estimate~$\hat{\bm{p}}_{1}^{(13)}$.\label{fig:Batlett_Traj}}
\end{figure}
%%%%%%%%%%%%%%%%%%%%%%%%%%%%%%%%%%%%%%%%%%
\section{Conclusions}\label{sec:conclusions}
A new, truly distributed MIMO channel sounding system was developed. The channel sounder was then used to perform measurements in an industry environment. The results show that distributing the antennas will achieve significant channel hardening and avoid deep fading dips due to small-scale and large-scale fading. We also investigated the stationarity regions, in which the \ac{wssus} assumptions held. This showed that the regions are quite small, with stationarity regions in the order of \num{2} m. We further showed that the RMS delay spread is in line with previous measurements conducted in similar settings and is around \SI{50}{\nano\second}; however, it varies between the distributed infrastructure antennas. Also, the Doppler spectral density was investigated by applying two super-resolution algorithms. We showed that in our data, both algorithms can find the dominant Doppler component.

Finally, we have highlighted the potential of positioning with D-MIMO in these environments.
Despite NLoS conditions, multipath propagation, and rich scattering in an industrial scenario, even a simple Bartlett beamformer can produce promising positioning results {with an \ac{mse} below \SI{20}{\centi\meter}} when paired with a suitable state-space filter. 
In the future, we will demonstrate a more elaborate estimator that outperforms our current solution. 
The initial results hint at possible centimeter-level positioning accuracy.

There are several directions for future work from here. Further investigations of channel characteristics are ongoing, where all the link combinations over the measured scenarios are classified as \ac{nlos} or \ac{los} and where all available data can contribute to the statistics of the channel. In addition, work investigating the performance of positioning capabilities in the more challenging \emph{{loop}} scenario is currently being carried out in parallel to further improvements of the positioning presented in this paper.
Another path is to investigate bi- or multistatic radar when the user is device free.

%
%\funding{This research was partially funded by Connected Systems at Lund University, the strategic research area ELLIIT, and the REINDEER project of the European Union’s
%Horizon 2020 research and innovation program under grant agreement no. 101013425.}

\appendix
\section*{Appendix A}\label{sec:bartlett-derivation}

Conventionally, the (empirical) Bartlett spectrum is defined as~\cite{Krim96ASP}

\begin{align}\label{eq:Bartlett-conv}
    \hat{P}(\theta) = \frac{\bm{a}(\theta)^\mathrm{H} \hat{\bm{R}} \bm{a}(\theta)}{\bm{a}^\mathrm{H}(\theta) \bm{a}(\theta)} \, ,
\end{align}
where $\bm{a}(\theta)$ is an array response parameterized on $\theta$ (often the angle of arrival of a spatial array response), and 

\begin{align}\label{eq:sample-cov-matrix}
    \hat{\bm{R}} = \frac{1}{N_x} \sum\limits_{t=1}^{N_x} \bm{x}(t) \bm{x}^\mathrm{H}(t)
\end{align}
is the sample covariance matrix of $N_x$ received signal vectors $\bm{x}$.

For multiple parameter estimation, {Equation} \eqref{eq:Bartlett-conv} can be used with a stacked vector of parameterized array responses. 
In the case of our Doppler-delay Bartlett beamformer, we thus choose $\bm{a}:=\vect\left( \bm{b}(\tau_n) \, \bm{c}^\mathrm{T}(\nu_n) \right) \in \mathbb{C}^{(\Nfrequency N_\nu \times 1)}$ and likewise, we stack the received signal matrix into a vector $\bm{x}:= \vect\left( \bm{H}_{\tilde{n}}\right)\in \mathbb{C}^{(\Nfrequency N_\nu \times 1)}$, where we use only $N_x=1$ of such vectors to compute~{Equation} \eqref{eq:sample-cov-matrix}. 
%Hence, we are left with 
Noting that $\bm{a}^\mathrm{H}\bm{a} = \lVert \bm{a} \rVert^2 \triangleq \Nfrequency N_\nu $, we formulate the Doppler-delay Bartlett spectrum for a single antenna $h$ as follows:

\begin{align}
    \hat{P}(\tau_n, \nu_n) &= \frac{1}{\Nfrequency N_\nu}
        \vect\left( \bm{b} \bm{c}^\mathrm{T} \right)^\mathrm{H}
        \vect\left( \bm{H}_{\tilde{n}}\right)
        \vect\left( \bm{H}_{\tilde{n}}\right)^\mathrm{H}
        \vect\left( \bm{b} \bm{c}^\mathrm{T} \right) \nonumber \\
        &= \frac{1}{\Nfrequency N_\nu}
        \left( \bm{c} \otimes \bm{b} \right)^\mathrm{H}
        \vect\left( \bm{H}_{\tilde{n}}\right)
        \vect\left( \bm{H}_{\tilde{n}}\right)^\mathrm{H}
        \left( \bm{c} \otimes \bm{b} \right) \nonumber \\
        &= \frac{1}{\Nfrequency N_\nu}
        \left( \bm{b}^\mathrm{H} \bm{H}_{\tilde{n}} \bm{c}^* \right)
        \left( \bm{b}^\mathrm{H} \bm{H}_{\tilde{n}} \bm{c}^* \right)^* \label{eq:vec-kron} \\
        &= \frac{\left| \bm{b}^\mathrm{H} \bm{H}_{\tilde{n}} \bm{c}^* \right|^2}{\Nfrequency N_\nu} \, , \label{eq:bartlett-derivation}
\end{align}
where we use the identity $\vect(\bm{A} \bm{B} \bm{C}) = \left( \bm{C}^\mathrm{T} \otimes \bm{A} \right) \vect(\bm{B})$ in~\eqref{eq:vec-kron} with $\otimes$ denoting the {Kronecker product.}

\bibliographystyle{IEEEtran}
\bibliography{IEEEabrv, ms}

\end{document}